\documentclass[11pt]{article}

\usepackage{graphicx}
\usepackage{amsmath}
\usepackage{amsfonts}
\usepackage{amssymb}
\usepackage{verbatim}

\makeatletter
\newdimen\normalarrayskip              % skip between lines
\newdimen\minarrayskip                 % minimal skip between lines
\normalarrayskip\baselineskip
\minarrayskip\jot
\newif\ifold             \oldtrue            \def\new{\oldfalse}
\def\arraymode{\ifold\relax\else\displaystyle\fi} % mode of array entries
\def\eqnumphantom{\phantom{(\theequation)}}     % right phantom in eqnarray
\def\@arrayskip{\ifold\baselineskip\z@\lineskip\z@
     \else
     \baselineskip\minarrayskip\lineskip2\minarrayskip\fi}
\def\@arrayclassz{\ifcase \@lastchclass \@acolampacol \or
\@ampacol \or \or \or \@addamp \or
   \@acolampacol \or \@firstampfalse \@acol \fi
\edef\@preamble{\@preamble
  \ifcase \@chnum
     \hfil$\relax\arraymode\@sharp$\hfil
     \or $\relax\arraymode\@sharp$\hfil
     \or \hfil$\relax\arraymode\@sharp$\fi}}
\def\@array[#1]#2{\setbox\@arstrutbox=\hbox{\vrule
     height\arraystretch \ht\strutbox
     depth\arraystretch \dp\strutbox
     width\z@}\@mkpream{#2}\edef\@preamble{\halign
\noexpand\@halignto
\bgroup \tabskip\z@ \@arstrut \@preamble \tabskip\z@ \cr}%
\let\@startpbox\@@startpbox \let\@endpbox\@@endpbox
  \if #1t\vtop \else \if#1b\vbox \else \vcenter \fi\fi
  \bgroup \let\par\relax
  \let\@sharp##\let\protect\relax
  \@arrayskip\@preamble}
\def\eqnarray{\stepcounter{equation}%
              \let\@currentlabel=\theequation
              \global\@eqnswtrue
              \global\@eqcnt\z@
              \tabskip\@centering
              \let\\=\@eqncr

 \halign to \displaywidth\bgroup
    \eqnumphantom\@eqnsel\hskip\@centering
    $\displaystyle \tabskip\z@ {##}$%
    \global\@eqcnt\@ne \hskip 2\arraycolsep
         $\displaystyle\arraymode{##}$\hfil
    \global\@eqcnt\tw@ \hskip 2\arraycolsep
         $\displaystyle\tabskip\z@{##}$\hfil
         \tabskip\@centering
    &{##}\tabskip\z@\cr}
\newfont{\hr}{msbm10}
\newfont{\ams}{msam10}
\voffset= - 1.2in
\hoffset= - 1.0in         % switch off for draft style
\def\beq{\begin{equation}}
\def\eeq{\end{equation}}
\def\ba{\beq\new\begin{array}{c}}
\def\ea{\end{array}\eeq}
\def\be{\ba}
\def\ee{\ea}

\def\N2{${\cal N}=2$}

\def\1N{${\cal N}=1$}
\def\4N{${\cal N}=4$}
\def\nn{\nonumber}

\def\p{\partial}

\def\p{\partial}
\def\tr{{\rm tr}\,}

\title{{\bf Complete Set of Cut-and-Join Operators
in Hurwitz-Kontsevich Theory}
\vspace{.5cm}}
\author{{\bf Andrei Mironov}\footnote{ {\small {\it
Lebedev Physics Institute} and {\it ITEP, Moscow, Russia}};
mironov@itep.ru; mironov@lpi.ru}, {\bf Alexei
Morozov}\thanks{{\small {\it ITEP, Moscow, Russia}};
morozov@itep.ru}\ \ and {\bf Sergey Natanzon}\thanks{{\small {\it Moscow
State University}, {\it ITEP} and
{\it Moscow Independent University, Moscow, Russia}};
natanzons@mail.ru} }

\begin{document}

\setcounter{footnote}{3}

\setcounter{tocdepth}{3}

\maketitle

\vspace{-7cm}

\begin{center}
\hfill FIAN/TD-06/09\\
\hfill ITEP/TH-16/09\\
\end{center}

\vspace{4.5cm}

\begin{abstract}
We define cut-and-join operator in Hurwitz theory
for merging of two branching points of arbitrary type.
These operators have two alternative descriptions:
(i) they have the $GL$ characters as eigenfunctions
and the symmetric-group characters as eigenvalues;
(ii) they can be represented as differential operators
of the $W$-type (in particular, acting on the time-variables in
the Hurwitz-Kontsevich tau-function).
The operators have the simplest form if expressed in terms
of the matrix Miwa-variables.
They form an important commutative associative algebra,
a Universal Hurwitz Algebra,
generalizing all group algebra centers of particular
symmetric groups which are used in description of the Universal
Hurwitz numbers of particular orders.
This algebra expresses arbitrary Hurwitz numbers as
values of a distinguished linear form on the linear space
of Young diagrams, evaluated at the product of all diagrams,
which characterize particular ramification points of the
covering.
\end{abstract}

\section{Introduction and summary}

\subsection{Hurwitz numbers and characters}

The Hurwitz numbers ${\rm Cov}_q(\Delta_1,\Delta_2,\ldots,\Delta_m)$
count a certainly weighted number of ramified $q$-fold coverings
of a Riemann sphere with fixed position of $m$ branching points of
the given types $\Delta_1,\ldots,\Delta_m$. The types are labeled by
ordered integer partitions of $q$, i.e. by the Young diagrams
$\Delta$ with $|\Delta|=q$ boxes. This seemingly formal problem
appears related to numerous directions of research in physics and
mathematics and attracts increasing attention in the literature: see
\cite{Hurfirst}-\cite{MShW} for some references. After an
accurate definition, see s.\ref{Hunu} below, the problem becomes
that of representation theory of symmetric groups and is reduced to
the celebrated formula \cite{Dijk} (our normalization of
$\varphi_R(\Delta)$ differs from that used in text-books by a factor):
\be
{\rm
Cov}_q(\Delta_1,\Delta_2,\ldots,\Delta_m) = \sum_{|R|=q} d_R^2
\varphi_R(\Delta_1)\varphi_R(\Delta_2)\ldots \varphi_R(\Delta_m)
\label{Frofor}
\ee
The r.h.s. is a sum over all representations
(Young diagrams) $R$ with $|R|=q$ and $\varphi_R(\Delta)$ are
expansion coefficients (in fact, these are proportional to the characters of
symmetric groups \cite{Mac}) of the $GL$ characters (Shur functions)\
$\chi_R(t)$ \cite{Mac} in time-variables $p_k=kt_k$:
\be
\chi_R(t) = \sum_{|\Delta|=|R|} d_R\varphi_R(\Delta) p\,(\Delta) =
\sum_{\Delta} d_R\varphi_R(\Delta)p\,(\Delta) \delta_{|\Delta|,|R|}
\label{varphichi}
\ee
For the integer partition $\Delta=[\mu_1,...]:\
\mu_1\geq \mu_2\geq\ldots \geq 0$ with $\sum_j \mu_j = |\Delta|$ a
monomial $p\,(\Delta) \equiv \prod_ip_{\mu_i}=\prod_j p_j^{\,m_j}$.
%%%If some adjacent $m_j$ coincide, one also needs
In what follows we also use a differently
normalized monomial: if $p\,(\Delta) = \prod_k p_k^{m_k}$, then
$\widetilde{p\,(\Delta)} \equiv \prod_k
\frac{1}{m_k!}\left(\frac{p_k}{k}\right)^{m_k} = \left(\prod_k m_k!
k^{m_k}\right)^{-1}\!p\,(\Delta)$. Further we use the same
definition of $\widetilde{Y(\Delta)}$ to define monomials for
arbitrary chains of variables $\{y_k\}$. The definitions of $\chi_R(t)$
and $d_R = \chi_R\Big(t_k=\delta_{k,1}\Big)$ are standard, see
s.\ref{chars} below.

One can extend the definition of $\varphi_R(\Delta)$ to
%%%$|\Delta| \neq |R|$
bigger diagrams $R$ with $|R| > |\Delta|$
in the following way:
\be
\varphi_R([\Delta,\underbrace{1,...,1}_{k}]) \equiv \left\{\begin{array}{ccc}
0 & {\rm for} &  |\Delta|+k>|R| \\
\varphi_R([\Delta,\underbrace{1,\ldots,1}_{|R|-|\Delta|}])
C^k_{|R|-|\Delta|} & {\rm for} &  |\Delta|+k\leq |R|
\end{array}\right.
\label{phiext}
\ee
Here $C^a_b={b!\over a!(b-a)!}$ are the binomial coefficients, and $\Delta$ is a Young diagram
that does not contain units, $1\notin \Delta$.
With this extension one can lift the requirement that all
$|\Delta_\alpha| = q$ in (\ref{Frofor}).

\subsection{Hurwitz partition functions}

There are two different ways to define generating functions of
Hurwitz numbers (see also \cite{Klemm} for
other generating functions related to partitions). First, $\varphi_R(\Delta)$ in
(\ref{Frofor}) can be contracted with $p(\Delta)$ and converted into
$\chi_R(t)$ with the help of (\ref{varphichi}). Second,
$\varphi_R(\Delta)$ can be exponentiated. This implies the following
definition of the Hurwitz partition function \cite{integ}: \be {\cal
Z}(t,t',t'',\ldots|\beta) \equiv \sum_R d_R^2 \frac{\chi_R(t)}{d_R}
\frac{\chi_R(t')}{d_R} \frac{\chi_R(t'')}{d_R}\ldots \exp
\left(\sum_\Delta\beta_\Delta\varphi_R(\Delta)\right) \label{calZ}
\ee where sum is now over all representations (Young diagrams) $R$
of arbitrary size $|R|$. Here $\beta$ is a set of constants
depending on Young diagrams. If only $\beta_{[2]}$ corresponding to
the diagrams $\Delta=[2]$ is not equal to zero, this reduces to the
generating of ${\cal N}$-Hurwitz numbers, in particular, \cite{GJV,OkToda}
\begin{equation}\label{Z1}
\begin{array}{lll}
{\cal N}=1: & Z(t|\beta) = \sum_R d_R\chi_R(t)e^{\beta_2\varphi_R([2])}
&\longrightarrow Z(t|0) = \sum_R d_R\chi_R(t) = e^{t_1},
\cr
\cr
{\cal N}=2:
& Z(t,\bar t|\beta) = \sum_R \chi_R(t)\chi_R(\bar t)
e^{\beta_2\varphi_R([2])}
&\longrightarrow Z(t,\bar t|0)
= \sum_R \chi_R(t)\chi_R(\bar t) = \exp\left(\sum_k kt_k\bar t_k\right)
\end{array}
\end{equation}
are KP and Toda-chain $\tau$-functions
in $t$ and $t,\bar t$
respectively \cite{KM2,OkToda,kaz,mmhk},
however integrability is violated
for ${\cal N}\geq 3$ \cite{integ}.
It is also violated by inclusion of higher $\beta_\Delta$
with $|\Delta|\geq 3$ \cite{OkOl,Lala,integ}:
to preserve it one should exponentiate the Casimir eigenvalues
$C_R(|\Delta|)$ \cite{KM2} instead of
$\varphi_R(\Delta)
\neq C_R(|\Delta|)$. Substitution of $\varphi_R$ by $C_R$ in the
definition of partition functions was nicknamed transition to the
complete cycles in \cite{OkOl,Lala}, ${\cal Z}$ is obtained from the
so defined $\tau$-function \cite{KM2} by action of sophisticated
operators ${\cal B}_\Delta$, see \cite{integ}.

The KP $\tau$-function $Z(t|\beta)$ is in fact related
\cite{giv,kaz,mmhk} by the equivalent-hierarchies technique \cite{eqhi}
to the Kontsevich $\tau$-function \cite{GKM} and,
following \cite{mmhk},
we call it and its further generalization (\ref{calZ})
Kontsevich-Hurwitz partition function.
This remarkable relation allows one to apply the
well-developed technique of matrix models \cite{GKM}-\cite{AMMIM}
to study of the Hurwitz numbers, and this paper develops further a
particular example \cite{MShW} of such application.

\subsection{General cut-and-join operators}

Alternatively one can introduce the $\beta$-deformations
in the partition function with the help of the cut-and-join operators
$\hat{\cal W}$, which are differential operators acting on the
time-variables $\{t_k\}$ (or, alternatively, on $\{t'_k\}$ or
$\{t''_k\}$) and have characters $\chi_R(t)$ as their eigenfunctions
and $\varphi_R(\Delta)$ as the corresponding eigenvalues:
\be
\boxed{
\hat{\cal W}(\Delta)\chi_R(t) = \varphi_R(\Delta) \chi_R(t)
}
\label{Wchi}
\ee
Then, as an immediate corollary of (\ref{Wchi}) and (\ref{calZ}),
\be
{\cal Z}(t,t',\ldots|\beta) = \exp\left(\sum_\Delta \beta_\Delta
\hat{\cal W}(\Delta)\right) {\cal Z}(t,t',\ldots|0)
\ee
In the simplest case of $\Delta=[2]$ we get the standard
cut-and-join operator \cite{GJV}
\be
\hat{\cal W}([2]) = \frac{1}{2}\sum_{a,b=1}^\infty
\left((a+b)p_ap_b\frac{\partial}{\partial p_{a+b}} +
abp_{a+b}\frac{\partial^2}{\partial p_a\partial p_b}\right)
= \frac{1}{2}\sum_{a,b=1}^\infty
\left(abt_at_b\frac{\partial}{\partial t_{a+b}} +
(a+b)t_{a+b}\frac{\partial^2}{\partial t_a\partial t_b}\right)
\label{W2}
\ee
which is the zero-mode generator
of $W_3$-algebra \cite{w3}:  $\hat{\cal W}([2]) = \hat W_0^{(3)}$.
The $W_3$-algebra is a part of the universal enveloping
algebra of $GL(\infty)$, and is a symmetry of the universal
Grassmannian \cite{Sato,UMS}, hence, the action of this operator preserves
KP-integrability \cite{Sato,DJKM,eqhi} and deforms Toda-integrability \cite{UT}
in a simple way \cite{KM2,OkToda}.

Operator (\ref{W2}) is conveniently rewritten \cite{MShW} in terms
of the matrix Miwa variable $X$ ($X$ was called $\psi$ in
\cite{MShW}), \be p_k = kt_k = \tr X^k, \ee where $X$ is an $N\times
N$ matrix, \be \hat {\cal W}([2]) = {1\over 2}\left(\tr \hat D^2 - N
\tr \hat D\right) ={1\over 2} \ :\tr \hat D^2: \label{W2thrD} \ee
Here we use the matrix operator $\hat D =
X\frac{\partial}{\partial\tilde X}$, involving transposed matrix
$\tilde X$ as usual in matrix models theory \cite{GKM}-\cite{AMMIM}
(hereafter, the repeated indices are implied to sum over):
\be\label{Dab} \hat D_{ab} \equiv X_{ac}\frac{\partial}{\partial X_{bc}} \ee
i.e. a family of operators $\left[\hat D_{ab},\hat D_{cd}\right]=
\hat D_{ad}\delta_{bc}-\hat D_{bc}\delta_{ad}$ acting on the algebra
of functions, generated by $X_{ab}$: $\hat D_{ab}X_{cd}\equiv
X_{ad}\delta_{bc}$. The normal ordering implies that the $X$-derivatives
do not act on $X$'s which stand between the two colons. It is
equivalent to taking a symbol of operator.

The goal of this paper is to claim that (\ref{W2thrD}) possesses
the immediate generalization to all other operators $\hat{\cal
W}(\Delta)$: \be \boxed{ \hat{\cal W}(\Delta) = \ :\widetilde{
D(\Delta)}: } \label{WDthrD} \ee where $\widetilde{ D(\Delta)}$ is
made from the operators $\hat D_k \equiv \tr \hat D^k$ in exactly
the same way as $\widetilde{p\,(\Delta)}$ is made from the
time-variables $p_k=kt_k$: $\widetilde{D\,(\Delta)} \equiv
\left(\prod_k m_k! k^{m_k}\right)^{-1}\!\hat D_k^{m_k}$. Note that
the operators $\hat D_{ab}$, (\ref{Dab}) realize the regular representation of
the algebra $gl$. The (commuting) Casimir operators that lie in the
universal enveloping algebra of $gl$ can be
realized as $\hat D_k$, the characters $\chi_R(t)$ of the group $GL$
being their eigenfunctions \cite{Helgason,ammops}. Since
all $\hat D_k$ commute among themselves, so do all the $\hat
D(\Delta)$ and $\hat{\cal W}(\Delta)$ (and their common system
of eigenfunctions is still formed by the characters). This allows
one to express ${\cal Z}(t|\beta)$ through the trivial
$\tau$-function $Z(t|0) = e^{t_1}$: \be \boxed{ {\cal Z}(t|\beta) =
\exp \left(\sum_\Delta\beta_\Delta :\widetilde{D(\Delta)}:\right)
Z(t|0) } \label{calZD} \ee Moreover, extra sets of time-variables
can be introduced with the help of the same operators, for example,
\be {\cal Z}(t,\bar t|\beta) \approx :e^{\sum_k \bar t_k \tr \hat
D^k}:\,{\cal Z}(t|\beta) \label{Ztt} \ee (For more accurate
formulation of what means $\approx$ in this equation see
s.\ref{Zttsec} below.) This opens a way to naturally incorporate
Hurwitz partition functions into the M-theory of matrix models
\cite{AMMIM}.

Beyond $\Delta=[2]$ the normal ordering makes even operators
$\hat{\cal W}([k])$ with a single-row Young diagram $\Delta=[k]$
non-linear combinations of the Casimir operators, this takes
$\hat{\cal W}(\Delta)$ out of the Universal Grassmannian \cite{Sato,UMS}
and leads to violation of integrability, observed in
\cite{OkOl,Lala,integ}.

Restricting the set $\{\beta_\Delta\}$ of $\beta$-variables in
(\ref{calZD}) to a single $\beta_{[2]}=\beta$ we obtain a representation
for the Kontsevich-Hurwitz $\tau$-function $Z(t|\beta)$, which was the
starting point in \cite{MShW} for the derivation of a promising
matrix model representation for this intriguing function. It is actually
expressed (in a yet badly-understood but clearly established way
\cite{giv,kaz,mmhk}) through the standard cubic Kontsevich
$\tau$-function \cite{GKM,AMM}.

The cut-and-join operators $\hat{\cal W}$ form a commutative associative
algebra, see s.\ref{Wal}: \be \boxed{ \hat{\cal
W}(\Delta_1)\hat{\cal W}(\Delta_2) = \sum_\Delta
C_{\Delta_1\Delta_2}^\Delta \hat{\cal W}(\Delta) } \label{CWW0} \ee
with the structure constants related to the triple Hurwitz numbers
$\hbox{Cov}(\Delta_1\Delta_2\Delta_3)$, see the next subsection 1.4. Accordingly these
$\hbox{Cov}(\Delta_1\Delta_2\Delta_3)$ can be alternatively studied in the
theory of \textit{dessins d'enfants} and Belyi functions
\cite{desBel}. At $|\Delta_1|=|\Delta_2|=|\Delta|$ these numbers are
the structure constants $c_{\Delta_1\Delta_2}^\Delta $
of the center of the group algebra of the symmetric group $S_{|\Delta|}$.

Eqs.(\ref{calZD}) and (\ref{Wal}) should possess an interesting
non-Abelian
generalization to the case of open Hurwitz numbers \cite{Na,AN,MMN3},
counting coverings of Riemann surfaces with boundaries,
which should be an open-string counterpart of the closed-string
formula (\ref{calZD}).

\subsection{Universal Hurwitz numbers and Universal Hurwitz algebra}

The structure constants $C^{\Delta}_{\Delta_1\Delta_2}$
allow one to introduce the Universal Hurwitz numbers,
defined for arbitrary sets of Young diagrams, not restricted
by the condition $|\Delta_1| = \ldots = |\Delta_m|$.

Consider the vector space $Y$ generated by all Young
diagrams. The correspondence
$\Delta\mapsto\hat{\mathcal{W}}(\Delta)$
generates a structure of commutative associative algebra on $Y$,
we denote the corresponding multiplication of Young diagrams
by $*$.
Consider a linear form $l:\ Y \rightarrow\mathbb{R}$, where
$l(\Delta)=\frac{1}{|\Delta|!}$ for $\Delta=[1,1,...,1]$ and
$l(\Delta)=0$ for all other Young diagrams. This definition
is motivated by eq.(\ref{pphi}) from the theory of characters,
see s.\ref{mf} below.
We call \textit{Hurwitz
number of $\Delta_1,\Delta_2,...,\Delta_m$} the number
\be
{\rm Cov}(\Delta_1,\Delta_2,...,\Delta_m)=
l(\Delta_1*\Delta_2*...*\Delta_m)
\ee
These generalized Hurwitz numbers coincide
with the classical ones for
$|\Delta_1|=|\Delta_2|=,...,=|\Delta_m|$,
when restricting the $*$-operation reproduces the composition $\circ$
of conjugation classes of permutations, considered in s.2.2.

The symmetric bilinear form $<\Delta_1,\Delta_2>\ =
l(\Delta_1*\Delta_2)$ is non-degenerate and invariant, \be
<\Delta_1*\Delta,\ \Delta_2>\ = \ <\Delta_1,\ \Delta_2*\Delta>\ \ \
\ \ \forall \Delta \ee as a corollary of commutativity and
associativity. Moreover, \be \sum_\Delta
C^{\Delta}_{\Delta_1\Delta_2} <\Delta,\Delta_3>\ =\
l(\Delta_1*\Delta_2*\Delta_3) \ee i.e. \be
C^{\Delta}_{\Delta_1\Delta_2} = \sum_{\Delta_3}G^{\Delta\Delta_3}
\ l(\Delta_1*\Delta_2*\Delta_3), \ee where $G^{\Delta_2\Delta_3}$ is
the inverse matrix of $G_{\Delta_1\Delta_2} = \
<\Delta_1,\Delta_2>$.

Finally, our Universal Hurwitz algebra of cut-and-join operators is freely generated
by a set of Casimir operators and actually coincides as a vector space with the center of the
universal enveloping algebra of $gl(\infty)$, see s.2.5 below and
\cite{integ,ammops} for more details.

\section{Comments}

\subsection{Hurwitz numbers and counting of coverings
\label{Hunu}}

A $q$-sheet covering $\Sigma$ of the Riemann surface $\Sigma_0$
is a projection $\pi:\ \Sigma\rightarrow\Sigma_0$,
where almost all the points of $\Sigma_0$ have exactly
$q$ pre-images.
The number of pre-images drops down at finitely many
singular (ramification) points $x_1,\ldots,x_m \in \Sigma_0$.
Actually, $\pi^{-1}(x_\alpha)$ is a collection of points
$y^{(\alpha)}_i\in \Sigma$, such that in the vicinity of each
$y^{(\alpha)}_i$ the projection $\pi$ looks like
\be
\pi:\ \ (x-x_\alpha) = (y-y^{(\alpha)}_i)^{\mu_i^{(\alpha)}}
\ee
Then with each singular point one associates an integer
partition of $q$, which can be ordered,
$\Delta_\alpha:\
\mu_1^{(\alpha)}\geq\mu_2^{(\alpha)}\geq\ldots\geq 0$,
i.e. is actually a Young diagram.
This diagram $\Delta_\alpha$ is named the \textit{type}
of ramification point $x_\alpha$.

If one picks up some non-singular point $x_*\in \Sigma_0$
and considers a closed path $C_*$ in $\Sigma_0$ which
begins and ends at $x_*$, then the $q$ pre-images of $x_*$
in $\Sigma$ get somehow permuted when one travels along $C_*$.
Thus with a path $C_*$ is associated a permutation of
$q$ variables, i.e. the covering defines a map from
the fundamental group $\pi_1(\Sigma_0,x_*)$ into the
symmetric (permutation) group $S_q$.
Changing $x_*$ amounts to the common conjugation of all
the permutations, associated with different contours,
and the covering itself is associated with the conjugated classes of maps
of $\pi_1(\Sigma_0,x_*)$ into $S_q$.
In fact, inverse is also almost true:
given such a map, one can reconstruct the covering.
Thus enumeration of ramified coverings becomes a pure group-theory problem
and gives the definition of Hurwitz numbers for the Riemann surface of arbitrary genus $g$:
\be
{\rm Cov}_q^g(\Delta_1,\ldots,\Delta_m) =
\sum \frac{1}{|{\rm Aut}(\pi)|}
\label{cov}
\ee
is the number
of its coverings $\pi$ with a fixed set of singular points
$x_1,\ldots,x_m$ of the types $\Delta_1,\ldots,\Delta_m$,
divided by the order of automorphism group. For the sake of brevity, we just
put Cov$_q^0=$Cov$_q$.
The sum in (\ref{cov}) is over all possible equivalence classes
of coverings and the equivalence is established by a bi-holomorphic map
$f:\ \Sigma \rightarrow \Sigma'$ such that
$\pi' = f\circ \pi$.

Since ${\rm Cov}_q(\Delta_1,\ldots,\Delta_m)$
is simultaneously a group-theory quantity,
it can be also expressed in terms of symmetric groups,
and this approach leads to formula
(\ref{Frofor}).
An extension to the surfaces $\Sigma_0$ of arbitrary genus follows from topological field theory
\cite{Atiah,DW,Dijk,Na}.
Somewhat non-trivial is generalization to $\Sigma_0$
with boundaries, see \cite{Na,AN}.

\subsection{Permutations, cycles and their compositions
\label{dpw}}

Cut-and-join operators come to the scene when one studies
merging of two ramification points $x_\alpha$ and
$x_\beta$ of the types $\Delta_\alpha$ and $\Delta_\beta$.
In result of such merging a single singular point emerges
at the place of two, but its type $\Delta$ is not
defined unambiguously by $\Delta_1$ and $\Delta_2$.
It depends on actual distribution of pre-images
$y_i^{(\alpha)}$ and $y_j^{(\beta)}$ between the sheets
of the covering, and this distribution is summed over
in the definition of Hurwitz numbers.

The Young diagram $\Delta$ labels the monodromy element of a critical
value and a conjugation class in the symmetric group $S_q$. When two
critical point merge, the resulting monodromy is a product of two
original monodromies.

Before we consider multiplication of classes, let us look
at multiplication of permutations.
Any permutation can be represented as a product of cycles.
For example, $S_3$ consists of six elements:
$$\{123\}\longrightarrow \{123\},\ \{132\},\ \{213\},\ \{321\}\ ,\{231\},\
\{312\},$$ which can be expressed through the cycles as
$$123,\ 1(23),\ (12)3,\ (13)2,\ (132),\ (123)$$  respectively.
The notation $(132)$ for a cycle means that
$1\rightarrow 3\rightarrow 2\rightarrow 1$.
For the sake of brevity we write $123$ instead of $(1)(2)(3)$.

The Young diagram $\Delta$ describes the conjugation class of elements of the group.
We denote with the same symbol $\Delta$ the element of the group algebra equal to the
sum of all elements of the conjugation class (with unit coefficients).

For instance, the Young diagrams with $3$ boxes label the three conjugation classes of
these permutations as follows:
$$
\Delta = [1,1,1] = 123;\ \ \
\Delta=[2,1] = 1(23),\ (12)3,\ (13)2; \ \ \
\Delta=[3] = (132),\ (123)
$$
The corresponding elements of the group algebra are
$$
\Delta = [1,1,1] = 123;\ \ \
\Delta=[2,1] = 1(23)\oplus (12)3\oplus (13)2; \ \ \
\Delta=[3] = (132)\oplus (123)
$$
It is convenient to define $||\Delta||$ as the number of
different permutations in the conjugation class $\Delta$,
e.g. $||3|| = 2$, $||2,1|| = 3$, $||1,1,1|| = 1$.

Similarly for $S_4$ there will be five conjugation classes and the corresponding
group algebra elements:
$$\Delta = [4] = (1234)\oplus (1243)\oplus (1324)\oplus (1342)
\oplus (1423)\oplus(1432),\ \ \ \ ||4|| = 3!=6,$$
$$\Delta = [3,1] = (123)4\oplus (124)3\oplus (132)4
\oplus (134)2\oplus (142)3\oplus (143)2\oplus 1(234)\oplus 1(243),
\ \ \ \ ||3,1|| = 8,$$
$$\Delta = [2,2] =  (12)(34)\oplus (13)(24)\oplus (14)(23),
\ \ \ \ ||2,2|| =3,$$
$$\Delta = [2,1,1]= (12)34\oplus (13)24\oplus (14)23\oplus 1(23)4
\oplus 1(24)3\oplus 12(34), \ \ \ \ ||2,1,1|| = 6,$$
and
$$\Delta = [1,1,1,1] = 1234, \ \ \ \ ||1,1,1,1|| = 1.$$

If we now consider merging of two ramification points, say, with
$\Delta=[2,1]$ and $\Delta'=[3]$, we need to see what happens when any of the three
permutations from the conjugation class $\Delta=[2,1]$ are multiplied by
any of the two from $\Delta'=[3]$. This is described by the $3\times 2$ table
\begin{equation}
[2,1]\circ [3] \ =\ \begin{array}{|c|c|}
\hline &\\
1(23) \circ (132) & 1(23) \circ (123) \\
&\\ \hline &\\
(12)3 \circ (132) & (12)3 \circ (123) \\
&\\ \hline &\\
(13)2 \circ (132) & (13)2 \circ (123) \\
&\\ \hline \end{array}
\ =\ \begin{array}{|c|c|}
\hline &\\
(13)2 & (12)3 \\
&\\ \hline &\\
1(23) & (13)2 \\
&\\ \hline &\\
(12)3 & 1(23) \\
&\\ \hline \end{array}\
=\ 2\cdot[2,1]
\label{21|3}
\end{equation}
or simply
\be
[2,1]\circ [3] \ =\
\Big(1(23)\oplus (12)3\oplus (13)2\Big)\circ
\Big((132)\oplus (123)\Big) =
2\cdot \Big(1(23)\oplus (12)3\oplus (13)2\Big)
=\ 2\cdot[2,1]
\ee
We denote the composition of permutations by $\circ$.
As usual, the second permutation acts first, for example,
$\{123\} \stackrel{(132)}{\longrightarrow} \{132\}
\stackrel{1(23)}{\longrightarrow} \{321\}$
and the result is the same as
$\{123\} \stackrel{(13)2}{\longrightarrow} \{321\}$.
Numbers in the notation of the permutation refer to
\textit{places}, not to \textit{elements}:
$(12)$ permutes the entries standing at the first
and the second place, not elements "$1$" and "$2$".

Having the composition of permutations, one can use the corresponding structure
constants $c_{\Delta\Delta'}^{\Delta''}$,
\be
\Delta\circ\Delta'=\sum_{\Delta''}c_{\Delta\Delta'}^{\Delta''}\Delta''
\ee
to define the cut-and-join operator by the following rule:
\be
\boxed{
\hat{\cal W}(\Delta) \widetilde{p\,(\Delta')}
= \sum_{\Delta''} c_{\Delta\Delta'}^{\Delta''} \widetilde{p\,(\Delta'')}
}
\label{DPW}
\ee
Eq.(\ref{21|3}) implies that the so defined operator
$\hat{\cal W}([2,1])$ contains an item
\be
\hat{\cal W}([2,1]) =
2\cdot\widetilde{p([2,1])}\frac{\p}{\p \widetilde {p([3])}} + \ldots
= 3p_1p_2\frac{\p}{\p p_3} + \ldots
\label{W21|3}
\ee
where dots  stand for the items that annihilate $p_3$.

Similarly, the composition table
\begin{equation}
[3]\circ [2,1] \ =\ \begin{array}{|c|c|c|}
\hline &&\\
(132) \circ 1(23) & (132) \circ  (12)3 & (123)  \circ (13)2 \\
&&\\ \hline &&\\
(123) \circ 1(23) & (123) \circ  (12)3 & (123)  \circ (13)2 \\
&&\\ \hline \end{array}
\ =\ \begin{array}{|c|c|c|}
\hline &&\\
(12)3 & (13)2 & 1(23) \\
&&\\ \hline &&\\
(13)2 & 1(23) & (12)3 \\
&&\\ \hline \end{array}\
=\ 2\cdot[2,1]
\label{3|21}
\end{equation}
implies that
\be
\hat{\cal W}([3]) = 2p_1p_2\frac{\p^2}{\p p_1\p p_2} + \ldots
\label{W3|21}
\ee
where this time dots stand for some terms from the annihilator
of $p_1p_2$.\footnote{
One can compare this formula with the full expression
in eq.(\ref{W3p}). Coefficient $2$ in (\ref{W3|21}) arises from
the second term in (\ref{W3p}) with $abcd = 1212, 1221, 2112, 2121$,
and only two out of these four terms contribute because of the
factor $(1-\delta_{ac}\delta_{bd})$.
}

The elements of the group algebra which correspond to the Young
diagrams, generate the center of the group algebra.
In our example
one can see that the r.h.s.'s of (\ref{3|21}) and (\ref{21|3}),
as implied by commutativity of the center.
%%% are the same, as a manifestation of this phenomenon.

In the same way one can analyze composition of any other pair of
conjugation classes and reconstruct all the entries in operators
$\hat{\cal W}(\Delta)$. In this way one can check that any
continuation of the first column in the Young diagram does not affect
the cut-and-join operator:
 \be \hat{\cal W}([\Delta,1,1,\ldots,1]) \cong
\hat{\cal W}(\Delta)\hspace{3cm} \hbox{if acting on a proper quantity}
\label{permid} \ee in accordance with
(\ref{phiext}), see sect.2.4.2 for details.
%%%The appropriately defined denominator in (\ref{calP}) is
%%%crucial for this property to occur.
We give just one more example, in a brief form:
\begin{equation}
[2,\underbrace{1,\ldots,1}_{q-2}] \circ
[3,\underbrace{1,\ldots,1}_{q-3}] \ =\
\begin{array}{|c|c|}
\hline & \\
(12)3456\ldots q \circ (123)456\ldots q & \ldots \\
\hline & \\
(13)2456\ldots q \circ (123)456\ldots q &  \\
\hline & \\
1(23)456\ldots q \circ (123)456\ldots q & \\
\hline & \\
(14)2356\ldots q \circ (123)456\ldots q &  \\
\hline & \\
\ldots & \\
\hline & \\
123(45)6\ldots q \circ (123)456\ldots q &  \\
\hline & \\
\ldots & \\
\hline
\end{array}  \ =\
\begin{array}{|c|c|}
\hline & \\
(13)2456\ldots q  & \ldots \\
\hline & \\
1(23)456\ldots q  &  \\
\hline & \\
(12)3456\ldots q  & \\
\hline\hline & \\
(1423)56\ldots q  &  \\
\hline & \\
\ldots & \\
\hline & \\
(123)(45)6\ldots q  &  \\
\hline & \\
\ldots & \\
\hline
\end{array}
\end{equation}
There are
$||3,\underbrace{1,\ldots,1}_{q-3}|| = 2C_q^3 = \frac{q(q-1)(q-2)}{3}$
columns and
$||2,\underbrace{1,\ldots,1}_{q-2}|| = C_q^2 = \frac{q(q-1)}{2}$
rows in the tables.
Clearly, each column of the second, resulting table
contains $3$ elements from the class
$[2,\underbrace{1,\ldots,1}_{q-2}]$ plus
$3(q-3)$ elements from the class
$[4,\underbrace{1,\ldots,1}_{q-4}]$ plus
$\frac{(q-3)(q-4)}{2}$ elements from the class
$[3,2,\underbrace{1,\ldots,1}_{q-5}]$.
Thus
\be
[2,1,\ldots,1]\circ \frac{[3,1,\ldots,1]}{||3,1,\ldots,1||}
= 3\cdot \frac{[2,1,\ldots,1]}{||2,1,\ldots,1||}
+ 3(q-3)\cdot \frac{[4,1,\ldots,1]}{||4,1,\ldots,1||}
+ \frac{(q-3)(q-4)}{2}\cdot \frac{[3,2,1,\ldots,1]}{||3,2,1,\ldots,1||}
\nn\ee
or
\be
[2,1,\ldots,1]\circ [3,1,\ldots,1]
= 2(q-2)\cdot [2,1,\ldots,1]
+ 4\cdot [4,1,\ldots,1]
+ [3,2,1,\ldots,1]
\label{2131}
\ee
Since in this example $\hat{\cal W}([2,\underbrace{1,\ldots,1}_{q-2}])$
acts on $p_3p_1^{q-3}$, we have:
\be
\hat{\cal W}([2,\underbrace{1,\ldots,1}_{q-2}])p_3p_1^{q-3} = \frac{1}{2}\left(
6p_1p_2\frac{\p}{\p p_3} + 6p_4\frac{\p^2}{\p p_1\p p_3}
+ p_2\frac{\p^2}{\p p_1^2} + \ldots\right)p_3p_1^{q-3}
\label{eW3}
\ee
We see that the coefficient in the term
$p_1p_2\frac{\p}{\p p_3}$ is the same as in (\ref{W21|3}),
in full accordance with (\ref{permid}).
%%%This example explicitly explains the choice of denominator in (\ref{calP}).
Both representations in (\ref{2131}) imply the same result for
$\hat{\cal W}([2,\underbrace{1,\ldots,1}_{q-2}])$ because
\vspace{-0.4cm}
$$
\widetilde {p\,(\Delta)} = \frac{||\Delta||}{|\Delta|!}p\,(\Delta)=
{p(\Delta )\over\hbox{Aut}(\Delta)}
$$
and both multiplication formulas can be used to extract cut-and-join operator
from (\ref{DPW}).

In general, for a composition of conjugation classes one has
\be
\boxed{ \Delta_1\circ \Delta_2 = \sum_{|\Delta| =
|\Delta_1|=|\Delta_2|} c_{\Delta_1\Delta_2}^\Delta \cdot \Delta }
\label{Ccc}
\ee
where small letter $c$ is used to stress that we deal with
the composition of permutations in the algebra $S_{|\Delta|}$, i.e.
$|\Delta| = |\Delta_1| = |\Delta_2|$.
Above examples demonstrate that even in this case the cut-and-join operator
is {\it not exactly}
$\sum_{|\Delta_1|=|\Delta_2|}
c_{\Delta\Delta_2}^{\Delta_1}\widetilde {p(\Delta_1)} {\partial}/{\partial
\widetilde{p(\Delta_2)}}$,\ the actual degree of differential operator which
satisfies (\ref{DPW}) can be much lower than implied by this formula.

In fact the constraint that $|\Delta'|=|\Delta|$ in (\ref{DPW}) can be easily
lifted: one can extend $\Delta$ to a diagram $[\Delta,1^{|\Delta'|-|\Delta|}]$
by adding a unit-height line of appropriate length and define
\be
\boxed{
\hat{\cal W}(\Delta) \widetilde{p\,(\Delta')}
= \widetilde{p}\,\Big([\Delta,1^{|\Delta'|-|\Delta|}]\circ\Delta'\Big)
= \sum_{|\Delta''|=|\Delta'|}
c_{[\Delta,1^{|\Delta'|-|\Delta|}]\,\Delta'}^{\Delta''} \widetilde{p\,(\Delta'')}
}\ \ \ \ {\rm for}\ 1\notin \Delta
\label{DPW1}
\ee
and
\be
\boxed{
\hat{\cal W}([\Delta,1^s]) \widetilde{p\,(\Delta')}
%= \widetilde{p\,\Big([\Delta,1^{|\Delta'|-|\Delta|}]\circ\Delta'\Big)}
= \sum_{|\Delta''|=|\Delta'|}
\frac{(|\Delta'|-|\Delta|)!}{s!(|\Delta'|-|\Delta|-s)!}\,
c_{[\Delta,1^{|\Delta'|-|\Delta|}]\,\Delta'}^{\Delta''} \widetilde{p\,(\Delta'')}
}\ \ \ \ {\rm for}\ 1\notin \Delta
\label{DPW2}
\ee

Thus cut-and-join operators can be defined as acting on the time-variables
of arbitrary level entirely in terms of the structure constants of the
universal symmetric algebra $S(\infty)$.
Eq.(\ref{WDthrD}), however, provides a much more explicit and transparent
alternative representation of these operators, which allows also to
extend the set of the $S(\infty)$ structure constants, by lifting the
remaining restriction $|\Delta''| = |\Delta'|$, which is still preserved
in (\ref{DPW1}) and (\ref{DPW2}). Extended structure constants
$C_{\Delta\Delta'}^{\Delta''}$ describe multiplication of the
universal operators, which are defined by either (\ref{DPW2}) or (\ref{WDthrD}).

\subsection{Composition of permutations and Feynman diagram technique
\label{dia}}

\begin{figure}[t]

\unitlength 0.8mm % = 2.845pt
\linethickness{0.4pt}
\ifx\plotpoint\undefined\newsavebox{\plotpoint}\fi % GNUPLOT compatibility
\hspace{-1cm}\begin{picture}(105.062,76.588)(0,0)
%\vector[middle](21.406,34.264)(19.176,35.453)
\put(20.291,34.859){\vector(-2,1){.07}}\multiput(21.406,34.264)(-.06194444,.03304167){36}{\line(-1,0){.06194444}}
%\end
%\vector[middle](21.926,35.379)(19.845,36.419)
\put(20.886,35.899){\vector(-2,1){.07}}\multiput(21.926,35.379)(-.0671452,.0335645){31}{\line(-1,0){.0671452}}
%\end
%\vector[middle](26.906,20.514)(26.237,18.656)
\put(26.571,19.585){\vector(-1,-3){.07}}\multiput(26.906,20.514)(-.03345,-.0929){20}{\line(0,-1){.0929}}
%\end
%\vector[middle](39.987,36.568)(41.771,37.906)
\put(40.879,37.237){\vector(4,3){.07}}\multiput(39.987,36.568)(.0445875,.03345){40}{\line(1,0){.0445875}}
%\end
%\vector[middle](40.656,35.602)(42.366,36.94)
\put(41.511,36.271){\vector(4,3){.07}}\multiput(40.656,35.602)(.0427375,.03345){40}{\line(1,0){.0427375}}
%\end
%\vector[middle](29.879,38.129)(29.284,40.73)
\put(29.582,39.43){\vector(-1,4){.07}}\multiput(29.879,38.129)(-.0330278,.1445278){18}{\line(0,1){.1445278}}
%\end
%\vector[middle](31.142,38.129)(30.696,40.805)
\put(30.919,39.467){\vector(-1,4){.07}}\multiput(31.142,38.129)(-.0318571,.1911429){14}{\line(0,1){.1911429}}
%\end
%\vector[middle](24.899,25.27)(23.041,23.784)
\put(23.97,24.527){\vector(-4,-3){.07}}\multiput(24.899,25.27)(-.0413,-.03303333){45}{\line(-1,0){.0413}}
%\end
%\vector[middle](24.156,26.46)(22.223,24.899)
\put(23.19,25.679){\vector(-4,-3){.07}}\multiput(24.156,26.46)(-.04111702,-.03321277){47}{\line(-1,0){.04111702}}
%\end
%\vector[middle](36.494,24.453)(38.203,22.223)
\put(37.349,23.338){\vector(3,-4){.07}}\multiput(36.494,24.453)(.03351961,-.04371569){51}{\line(0,-1){.04371569}}
%\end
%\vector[middle](37.46,25.345)(39.095,23.115)
\put(38.278,24.23){\vector(3,-4){.07}}\multiput(37.46,25.345)(.03336735,-.0455){49}{\line(0,-1){.0455}}
%\end
%\vector[middle](81.238,52.102)(78.339,52.028)
\put(79.789,52.065){\vector(-1,0){.07}}\put(81.238,52.102){\line(-1,0){2.8985}}
%\end
%\vector[middle](81.312,50.838)(78.488,50.838)
\put(79.9,50.838){\vector(-1,0){.07}}\put(81.312,50.838){\line(-1,0){2.8245}}
%\end
%\vector[middle](51.433,51.879)(54.852,51.953)
\put(53.143,51.916){\vector(1,0){.07}}\put(51.433,51.879){\line(1,0){3.419}}
%\end
%\vector[middle](51.433,50.541)(54.406,50.541)
\put(52.92,50.541){\vector(1,0){.07}}\put(51.433,50.541){\line(1,0){2.973}}
%\end
%\vector[middle](58.048,38.054)(59.832,40.21)
\put(58.94,39.132){\vector(3,4){.07}}\multiput(58.048,38.054)(.03365094,.04066981){53}{\line(0,1){.04066981}}
%\end
%\vector[middle](59.089,37.46)(60.575,39.541)
\put(59.832,38.5){\vector(3,4){.07}}\multiput(59.089,37.46)(.03303333,.04624444){45}{\line(0,1){.04624444}}
%\end
%\vector[middle](72.467,36.94)(70.758,39.615)
\put(71.613,38.278){\vector(-2,3){.07}}\multiput(72.467,36.94)(-.03351961,.05246078){51}{\line(0,1){.05246078}}
%\end
%\vector[middle](73.508,37.46)(71.947,40.061)
\put(72.728,38.761){\vector(-2,3){.07}}\multiput(73.508,37.46)(-.03320213,.05535106){47}{\line(0,1){.05535106}}
%\end
%\vector[middle](57.147,62.306)(59.268,59.919)
\put(58.208,61.113){\vector(1,-1){.07}}\multiput(57.147,62.306)(.0336746,-.03788095){63}{\line(0,-1){.03788095}}
%\end
%\vector[middle](58.296,63.013)(60.064,60.803)
\put(59.18,61.908){\vector(3,-4){.07}}\multiput(58.296,63.013)(.03334906,-.04169811){53}{\line(0,-1){.04169811}}
%\end
%\vector[middle](73.587,63.631)(71.996,60.626)
\put(72.792,62.129){\vector(-1,-2){.07}}\multiput(73.587,63.631)(-.03314583,-.06260417){48}{\line(0,-1){.06260417}}
%\end
%\vector[middle](74.648,62.836)(73.234,60.361)
\put(73.941,61.599){\vector(-1,-2){.07}}\multiput(74.648,62.836)(-.03366667,-.05891667){42}{\line(0,-1){.05891667}}
%\end
%\vector[middle](50.871,58.858)(52.993,57.533)
\put(51.932,58.196){\vector(3,-2){.07}}\multiput(50.871,58.858)(.0530375,-.0331375){40}{\line(1,0){.0530375}}
%\end
%\vector[middle](51.49,60.096)(53.611,58.593)
\put(52.551,59.345){\vector(4,-3){.07}}\multiput(51.49,60.096)(.04713333,-.03338889){45}{\line(1,0){.04713333}}
%\end
%\vector[middle](65.455,68.051)(65.632,65.664)
\put(65.544,66.858){\vector(0,-1){.07}}\multiput(65.455,68.051)(.0295,-.39775){6}{\line(0,-1){.39775}}
%\end
%\vector[middle](66.693,67.874)(66.781,65.664)
\put(66.737,66.769){\vector(0,-1){.07}}\put(66.693,67.874){\line(0,-1){2.21}}
%\end
%\vector[middle](80.393,60.273)(78.714,58.858)
\put(79.554,59.566){\vector(-4,-3){.07}}\multiput(80.393,60.273)(-.0399881,-.03367857){42}{\line(-1,0){.0399881}}
%\end
%\vector[middle](81.1,59.124)(79.509,57.886)
\put(80.305,58.505){\vector(-4,-3){.07}}\multiput(81.1,59.124)(-.043,-.03344595){37}{\line(-1,0){.043}}
%\end
%\vector[middle](81.277,42.33)(79.509,43.39)
\put(80.393,42.86){\vector(-3,2){.07}}\multiput(81.277,42.33)(-.0552344,.0331406){32}{\line(-1,0){.0552344}}
%\end
%\vector[middle](80.57,41.181)(78.802,42.33)
\put(79.686,41.755){\vector(-3,2){.07}}\multiput(80.57,41.181)(-.0505,.03282857){35}{\line(-1,0){.0505}}
%\end
%\vector[middle](67.135,33.226)(67.135,35.612)
\put(67.135,34.419){\vector(0,1){.07}}\put(67.135,33.226){\line(0,1){2.3865}}
%\end
%\vector[middle](65.809,33.314)(65.809,35.612)
\put(65.809,34.463){\vector(0,1){.07}}\put(65.809,33.314){\line(0,1){2.298}}
%\end
%\vector[middle](51.402,41.446)(53.346,42.772)
\put(52.374,42.109){\vector(3,2){.07}}\multiput(51.402,41.446)(.0486125,.03315){40}{\line(1,0){.0486125}}
%\end
%\vector[middle](52.286,40.385)(54.23,41.799)
\put(53.258,41.092){\vector(4,3){.07}}\multiput(52.286,40.385)(.04629762,.03366667){42}{\line(1,0){.04629762}}
%\end
\qbezier(21.394,34.242)(19.582,29.823)(22.189,24.873)
\qbezier(22.985,23.812)(24.023,21.934)(26.918,20.498)
\qbezier(27.935,20.056)(33.923,18.487)(38.232,22.133)
\qbezier(39.248,23.017)(44.021,28.762)(40.663,35.568)
\qbezier(40.05,36.54)(35.898,41.322)(30.695,40.849)
\qbezier(29.276,40.744)(24.625,39.851)(21.971,35.383)
\qbezier(54.135,41.638)(55.423,39.956)(58.077,38.064)
\qbezier(59.128,37.328)(61.572,35.83)(65.803,35.489)
\qbezier(67.169,35.541)(69.981,35.699)(72.477,36.908)
\qbezier(73.559,37.378)(76.997,39.316)(78.934,42.253)
\qbezier(79.559,43.378)(81.403,46.628)(81.372,50.878)
\qbezier(81.238,52.176)(80.94,55.447)(79.454,57.825)
\qbezier(78.714,58.947)(76.681,61.731)(74.648,62.924)
\qbezier(73.587,63.631)(70.449,65.443)(66.781,65.664)
\qbezier(65.632,65.664)(60.815,65.311)(58.296,63.189)
\qbezier(57.235,62.394)(55.158,60.891)(53.611,58.682)
\qbezier(52.92,57.602)(51.582,55.075)(51.433,51.953)
\qbezier(51.433,50.541)(51.21,46.453)(53.366,42.663)
\qbezier(42.298,36.761)(47.91,40.783)(44.154,50.638)
\qbezier(41.591,37.822)(46.098,41.446)(42.828,50.02)
%\qbezvec[middle](44.242,50.462)(40.751,63.764)(50.871,58.858)
\put(44.154,59.212){\vector(1,2){.07}}\qbezier(44.242,50.462)(40.751,63.764)(50.871,58.858)
%\end
%\qbezvec[middle](42.828,49.843)(38.674,66.062)(51.49,60.184)
\put(42.917,60.538){\vector(2,3){.07}}\qbezier(42.828,49.843)(38.674,66.062)(51.49,60.184)
%\end
%\qbezvec[middle](19.869,36.329)(9.936,41.322)(19.553,57.457)
\put(14.823,44.108){\vector(0,1){.07}}\qbezier(19.869,36.329)(9.936,41.322)(19.553,57.457)
%\end
%\qbezvec[middle](19.343,35.383)(7.676,40.586)(18.292,58.193)
\put(13.247,43.687){\vector(0,1){.07}}\qbezier(19.343,35.383)(7.676,40.586)(18.292,58.193)
%\end
%\qbezvec[middle](19.553,57.457)(28.278,72.225)(56.343,73.749)
\put(33.113,68.914){\vector(2,1){.07}}\qbezier(19.553,57.457)(28.278,72.225)(56.343,73.749)
%\end
%\qbezvec[middle](18.292,58.193)(27.121,74.012)(56.132,75.326)
\put(32.167,70.386){\vector(2,1){.07}}\qbezier(18.292,58.193)(27.121,74.012)(56.132,75.326)
%\end
\qbezier(87.456,42.846)(84.828,38.432)(80.518,41.165)
\qbezier(85.879,43.582)(84.565,40.639)(81.359,42.321)
\qbezier(26.184,18.628)(22.184,6.534)(45.934,7.816)
\qbezier(27.372,18.316)(24.028,7.972)(45.559,9.253)
%\vector[middle](28.059,20.003)(27.372,18.253)
\put(27.716,19.128){\vector(-1,-3){.07}}\multiput(28.059,20.003)(-.0327381,-.0833333){21}{\line(0,-1){.0833333}}
%\end
%\qbezvec[middle](45.516,9.211)(70.953,9.894)(88.822,27.185)
\put(69.061,14.046){\vector(3,1){.07}}\qbezier(45.516,9.211)(70.953,9.894)(88.822,27.185)
%\end
%\qbezvec[middle](45.831,7.844)(71.847,8.265)(90.084,26.344)
\put(69.902,12.68){\vector(3,1){.07}}\qbezier(45.831,7.844)(71.847,8.265)(90.084,26.344)
%\end
%\qbezvec[middle](90.084,26.239)(105.062,39.483)(98.387,58.193)
\put(99.649,40.849){\vector(1,4){.07}}\qbezier(90.084,26.239)(105.062,39.483)(98.387,58.193)
%\end
%\qbezvec[middle](88.822,27.08)(102.855,39.903)(96.916,57.352)
\put(97.862,41.06){\vector(1,3){.07}}\qbezier(88.822,27.08)(102.855,39.903)(96.916,57.352)
%\end
%\qbezvec[middle](98.493,58.088)(93.552,70.386)(80.624,60.4)
\put(91.555,64.815){\vector(-4,1){.07}}\qbezier(98.493,58.088)(93.552,70.386)(80.624,60.4)
%\end
%\qbezvec[middle](97.01,57.267)(93.121,68.183)(81.1,59.124)
\put(91.088,63.19){\vector(-4,1){.07}}\qbezier(97.01,57.267)(93.121,68.183)(81.1,59.124)
%\end
\put(26.014,28.095){\makebox(0,0)[cc]{1}}
\put(17.838,33.595){\makebox(0,0)[cc]{1}}
\put(32.703,37.163){\makebox(0,0)[cc]{2}}
\put(42.96,34.784){\makebox(0,0)[cc]{2}}
\put(37.311,27.649){\makebox(0,0)[cc]{3}}
\put(23.933,19.473){\makebox(0,0)[cc]{3}}
\put(31.663,30.325){\oval(6.54,6.838)[r]}
\put(66.56,50.506){\oval(6.54,6.838)[r]}
\put(31.872,33.691){\vector(-1,0){.625}}
\put(66.769,53.872){\vector(-1,0){.625}}
%\qbezvec[middle](56.343,73.645)(78.785,75.011)(85.039,63.344)
\put(74.738,71.753){\vector(3,-1){.07}}\qbezier(56.343,73.645)(78.785,75.011)(85.039,63.344)
%\end
%\qbezvec[middle](56.343,75.327)(79.993,76.588)(86.405,63.974)
\put(75.684,73.119){\vector(3,-1){.07}}\qbezier(56.343,75.327)(79.993,76.588)(86.405,63.974)
%\end
%\qbezvec[middle](85.785,61.953)(91,53.467)(85.962,43.568)
\put(88.437,53.114){\vector(0,-1){.07}}\qbezier(85.785,61.953)(91,53.467)(85.962,43.568)
%\end
%\qbezvec[middle](87.2,62.571)(92.547,54.042)(87.465,42.861)
\put(89.94,53.379){\vector(0,-1){.07}}\qbezier(87.2,62.571)(92.547,54.042)(87.465,42.861)
%\end
\end{picture}

\vspace{-6cm}

\unitlength 0.85mm % = 2.845pt
\linethickness{0.4pt}
\ifx\plotpoint\undefined\newsavebox{\plotpoint}\fi % GNUPLOT compatibility
\hspace{8cm}\begin{picture}(92.972,76.049)(0,0)
%\vector[middle](20.562,33.567)(18.333,34.757)
\put(19.447,34.162){\vector(-2,1){.07}}\multiput(20.562,33.567)(-.06193787,.03303353){36}{\line(-1,0){.06193787}}
%\end
%\vector[middle](21.083,34.682)(19.001,35.723)
\put(20.042,35.203){\vector(-2,1){.07}}\multiput(21.083,34.682)(-.0671327,.0335663){31}{\line(-1,0){.0671327}}
%\end
%\vector[middle](26.062,19.817)(25.393,17.959)
\put(25.728,18.888){\vector(-1,-3){.07}}\multiput(26.062,19.817)(-.0334465,-.0929068){20}{\line(0,-1){.0929068}}
%\end
%\vector[middle](39.144,35.871)(40.927,37.209)
\put(40.036,36.54){\vector(4,3){.07}}\multiput(39.144,35.871)(.04459527,.03344645){40}{\line(1,0){.04459527}}
%\end
%\vector[middle](39.813,34.905)(41.522,36.243)
\put(40.667,35.574){\vector(4,3){.07}}\multiput(39.813,34.905)(.04273713,.03344645){40}{\line(1,0){.04273713}}
%\end
%\vector[middle](29.035,37.432)(28.441,40.034)
\put(28.738,38.733){\vector(-1,4){.07}}\multiput(29.035,37.432)(-.0330335,.1445217){18}{\line(0,1){.1445217}}
%\end
%\vector[middle](30.299,37.432)(29.853,40.108)
\put(30.076,38.77){\vector(-1,4){.07}}\multiput(30.299,37.432)(-.0318538,.1911226){14}{\line(0,1){.1911226}}
%\end
%\vector[middle](24.056,24.574)(22.197,23.088)
\put(23.126,23.831){\vector(-4,-3){.07}}\multiput(24.056,24.574)(-.04129191,-.03303353){45}{\line(-1,0){.04129191}}
%\end
%\vector[middle](23.312,25.763)(21.38,24.202)
\put(22.346,24.983){\vector(-4,-3){.07}}\multiput(23.312,25.763)(-.0411162,-.03320924){47}{\line(-1,0){.0411162}}
%\end
%\vector[middle](35.65,23.756)(37.36,21.527)
\put(36.505,22.642){\vector(3,-4){.07}}\multiput(35.65,23.756)(.03351932,-.04372085){51}{\line(0,-1){.04372085}}
%\end
%\vector[middle](36.617,24.648)(38.252,22.419)
\put(37.434,23.533){\vector(3,-4){.07}}\multiput(36.617,24.648)(.03337061,-.04550537){49}{\line(0,-1){.04550537}}
%\end
%\vector[middle](80.394,51.406)(77.496,51.331)
\put(78.945,51.368){\vector(-1,0){.07}}\put(80.394,51.406){\line(-1,0){2.8987}}
%\end
%\vector[middle](80.469,50.142)(77.644,50.142)
\put(79.056,50.142){\vector(-1,0){.07}}\put(80.469,50.142){\line(-1,0){2.8244}}
%\end
%\vector[middle](50.59,51.183)(54.009,51.257)
\put(52.299,51.22){\vector(1,0){.07}}\put(50.59,51.183){\line(1,0){3.419}}
%\end
%\vector[middle](50.59,49.845)(53.563,49.845)
\put(52.076,49.845){\vector(1,0){.07}}\put(50.59,49.845){\line(1,0){2.973}}
%\end
%\vector[middle](57.205,37.358)(58.989,39.513)
\put(58.097,38.436){\vector(3,4){.07}}\multiput(57.205,37.358)(.03365681,.04066864){53}{\line(0,1){.04066864}}
%\end
%\vector[middle](58.245,36.763)(59.732,38.845)
\put(58.989,37.804){\vector(3,4){.07}}\multiput(58.245,36.763)(.03303353,.04624694){45}{\line(0,1){.04624694}}
%\end
%\vector[middle](71.624,36.243)(69.914,38.919)
\put(70.769,37.581){\vector(-2,3){.07}}\multiput(71.624,36.243)(-.03351932,.05246502){51}{\line(0,1){.05246502}}
%\end
%\vector[middle](72.664,36.763)(71.104,39.365)
\put(71.884,38.064){\vector(-2,3){.07}}\multiput(72.664,36.763)(-.03320924,.05534874){47}{\line(0,1){.05534874}}
%\end
%\vector[middle](56.303,61.609)(58.425,59.223)
\put(57.364,60.416){\vector(1,-1){.07}}\multiput(56.303,61.609)(.03367175,-.03788072){63}{\line(0,-1){.03788072}}
%\end
%\vector[middle](57.453,62.316)(59.22,60.106)
\put(58.336,61.211){\vector(3,-4){.07}}\multiput(57.453,62.316)(.03335409,-.04169262){53}{\line(0,-1){.04169262}}
%\end
%\vector[middle](72.744,62.935)(71.153,59.93)
\put(71.948,61.432){\vector(-1,-2){.07}}\multiput(72.744,62.935)(-.03314563,-.06260841){48}{\line(0,-1){.06260841}}
%\end
%\vector[middle](73.804,62.139)(72.39,59.665)
\put(73.097,60.902){\vector(-1,-2){.07}}\multiput(73.804,62.139)(-.03367175,-.05892557){42}{\line(0,-1){.05892557}}
%\end
%\vector[middle](50.028,58.162)(52.149,56.836)
\put(51.089,57.499){\vector(3,-2){.07}}\multiput(50.028,58.162)(.05303301,-.03314563){40}{\line(1,0){.05303301}}
%\end
%\vector[middle](50.647,59.399)(52.768,57.897)
\put(51.707,58.648){\vector(4,-3){.07}}\multiput(50.647,59.399)(.04714045,-.03339115){45}{\line(1,0){.04714045}}
%\end
%\vector[middle](64.612,67.354)(64.789,64.968)
\put(64.7,66.161){\vector(0,-1){.07}}\multiput(64.612,67.354)(.029463,-.397748){6}{\line(0,-1){.397748}}
%\end
%\vector[middle](65.849,67.178)(65.938,64.968)
\put(65.894,66.073){\vector(0,-1){.07}}\put(65.849,67.178){\line(0,-1){2.2097}}
%\end
%\vector[middle](79.55,59.576)(77.87,58.162)
\put(78.71,58.869){\vector(-4,-3){.07}}\multiput(79.55,59.576)(-.0399852,-.03367175){42}{\line(-1,0){.0399852}}
%\end
%\vector[middle](80.257,58.427)(78.666,57.19)
\put(79.461,57.808){\vector(-4,-3){.07}}\multiput(80.257,58.427)(-.04299974,-.03344424){37}{\line(-1,0){.04299974}}
%\end
%\vector[middle](80.434,41.633)(78.666,42.694)
\put(79.55,42.164){\vector(-3,2){.07}}\multiput(80.434,41.633)(-.0552427,.0331456){32}{\line(-1,0){.0552427}}
%\end
%\vector[middle](79.726,40.484)(77.959,41.633)
\put(78.843,41.059){\vector(-3,2){.07}}\multiput(79.726,40.484)(-.05050763,.03282996){35}{\line(-1,0){.05050763}}
%\end
%\vector[middle](66.291,32.529)(66.291,34.916)
\put(66.291,33.723){\vector(0,1){.07}}\put(66.291,32.529){\line(0,1){2.3865}}
%\end
%\vector[middle](64.966,32.618)(64.966,34.916)
\put(64.966,33.767){\vector(0,1){.07}}\put(64.966,32.618){\line(0,1){2.2981}}
%\end
%\vector[middle](50.558,40.749)(52.503,42.075)
\put(51.531,41.412){\vector(3,2){.07}}\multiput(50.558,40.749)(.04861359,.03314563){40}{\line(1,0){.04861359}}
%\end
%\vector[middle](51.442,39.689)(53.387,41.103)
\put(52.414,40.396){\vector(4,3){.07}}\multiput(51.442,39.689)(.04629866,.03367175){42}{\line(1,0){.04629866}}
%\end
\qbezier(20.55,33.546)(18.738,29.126)(21.346,24.177)
\qbezier(22.141,23.116)(23.18,21.238)(26.075,19.801)
\qbezier(27.091,19.359)(33.079,17.791)(37.388,21.437)
\qbezier(38.405,22.32)(43.178,28.066)(39.819,34.872)
\qbezier(39.207,35.843)(35.055,40.626)(29.852,40.153)
\qbezier(28.433,40.048)(23.782,39.154)(21.128,34.687)
\qbezier(53.292,40.941)(54.579,39.259)(57.234,37.367)
\qbezier(58.285,36.632)(60.728,35.134)(64.959,34.792)
\qbezier(66.326,34.845)(69.137,35.002)(71.634,36.211)
\qbezier(72.716,36.682)(76.153,38.619)(78.091,41.557)
\qbezier(78.716,42.682)(80.56,45.932)(80.528,50.182)
\qbezier(80.394,51.48)(80.097,54.75)(78.61,57.129)
\qbezier(77.87,58.25)(75.837,61.035)(73.804,62.228)
\qbezier(72.744,62.935)(69.606,64.747)(65.938,64.968)
\qbezier(64.789,64.968)(59.972,64.614)(57.453,62.493)
\qbezier(56.392,61.697)(54.315,60.195)(52.768,57.985)
\qbezier(52.076,56.906)(50.738,54.379)(50.59,51.257)
\qbezier(50.59,49.845)(50.367,45.757)(52.522,41.966)
\qbezier(41.454,36.065)(47.067,40.087)(43.31,49.942)
\qbezier(40.747,37.126)(45.255,40.749)(41.985,49.323)
%\qbezvec[middle](43.399,49.765)(39.907,63.068)(50.028,58.162)
\put(43.31,58.516){\vector(1,2){.07}}\qbezier(43.399,49.765)(39.907,63.068)(50.028,58.162)
%\end
%\qbezvec[middle](41.985,49.146)(37.83,65.366)(50.647,59.488)
\put(42.073,59.841){\vector(2,3){.07}}\qbezier(41.985,49.146)(37.83,65.366)(50.647,59.488)
%\end
%\qbezvec[middle](19.025,35.633)(9.092,40.626)(18.71,56.761)
\put(13.98,43.411){\vector(0,1){.07}}\qbezier(19.025,35.633)(9.092,40.626)(18.71,56.761)
%\end
%\qbezvec[middle](18.5,34.687)(6.832,39.89)(17.449,57.496)
\put(12.403,42.991){\vector(0,1){.07}}\qbezier(18.5,34.687)(6.832,39.89)(17.449,57.496)
%\end
%\qbezvec[middle](18.71,56.761)(27.434,71.529)(55.499,73.053)
\put(32.269,68.218){\vector(2,1){.07}}\qbezier(18.71,56.761)(27.434,71.529)(55.499,73.053)
%\end
%\qbezvec[middle](17.449,57.496)(26.278,73.316)(55.289,74.63)
\put(31.323,69.689){\vector(2,1){.07}}\qbezier(17.449,57.496)(26.278,73.316)(55.289,74.63)
%\end
%\qbezvec[middle](55.394,73.053)(75.103,74.314)(82.828,64.644)
\put(72.107,71.581){\vector(4,-1){.07}}\qbezier(55.394,73.053)(75.103,74.314)(82.828,64.644)
%\end
%\qbezvec[middle](55.394,74.63)(76.154,76.049)(84.09,65.485)
\put(72.948,73.053){\vector(3,-1){.07}}\qbezier(55.394,74.63)(76.154,76.049)(84.09,65.485)
%\end
%\qbezvec[middle](83.985,65.59)(92.972,53.344)(86.612,42.15)
\put(89.135,53.607){\vector(1,-4){.07}}\qbezier(83.985,65.59)(92.972,53.344)(86.612,42.15)
%\end
%\qbezvec[middle](82.933,64.434)(90.869,53.292)(85.141,42.991)
\put(87.453,53.502){\vector(1,-4){.07}}\qbezier(82.933,64.434)(90.869,53.292)(85.141,42.991)
%\end
\qbezier(86.612,42.15)(83.985,37.735)(79.675,40.468)
\qbezier(85.036,42.886)(83.722,39.943)(80.516,41.624)
\qbezier(25.341,17.932)(21.341,5.838)(45.091,7.119)
\qbezier(26.528,17.619)(23.185,7.276)(44.716,8.557)
%\vector[middle](27.216,19.307)(26.528,17.557)
\put(26.872,18.432){\vector(-1,-3){.07}}\multiput(27.216,19.307)(-.0327381,-.0833333){21}{\line(0,-1){.0833333}}
%\end
\put(25.17,27.398){\makebox(0,0)[cc]{1}}
\put(16.995,32.898){\makebox(0,0)[cc]{1}}
\put(31.86,36.466){\makebox(0,0)[cc]{2}}
\put(42.117,34.088){\makebox(0,0)[cc]{2}}
\put(36.468,26.952){\makebox(0,0)[cc]{3}}
\put(23.089,18.777){\makebox(0,0)[cc]{3}}
\put(30.819,29.628){\oval(6.541,6.838)[r]}
\put(65.716,49.81){\oval(6.541,6.838)[r]}
\put(31.028,32.994){\vector(-1,0){.625}}
\put(65.925,53.176){\vector(-1,0){.625}}
%\qbezvec[middle](44.673,8.619)(64.802,11.825)(64.959,32.48)
\put(59.809,16.187){\vector(1,1){.07}}\qbezier(44.673,8.619)(64.802,11.825)(64.959,32.48)
%\end
%\qbezvec[middle](44.988,7.043)(66.273,10.248)(66.326,32.585)
\put(60.965,15.031){\vector(1,1){.07}}\qbezier(44.988,7.043)(66.273,10.248)(66.326,32.585)
%\end
\end{picture}
\caption{\footnotesize{Composition of two permutations: of the cycle
(123) with the order 6 cycle. At the same time, these
Feynman diagrams contribute to multiplication
of the normal ordered matrix differential operators $\hat{\cal W}([3])$ and
$\hat{\cal W}([6])$.}
\label{fig}}
\end{figure}
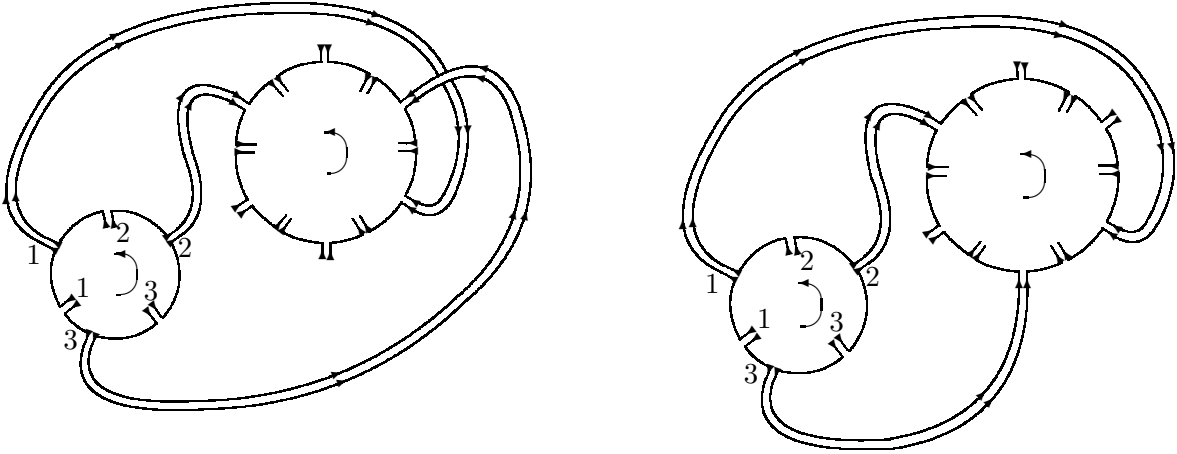

Composition of permutations can be conveniently calculated
with the help of a simple Feynman diagram technique.
This, on one hand, literally reflects the geometric definition of the Hurwitz numbers
and, on the other hand, is equivalent to the description through differential operators.
Represent a cycle $(132)$ of length $3$ by an oriented circle at the l.h.s.
of Fig.\ref{fig} and a cycle of length $6$ by another oriented
circle at its r.h.s. The composition itself is represented
by lines connecting all outgoing lines of the left circle
with arbitrarily chosen $3$ incoming lines of the right circle.
As a result, the new cycles are formed: just one of length $6$
for connecting lines as at the left figure and three of the lengths $1$, $2$
and $3$ if one of the connecting lines is changed as at the right figure.

In this picture we deal with the situation
of the type $(123)\circ (123456)$, when the first cycle
is a subset of the second one. One should only keep in mind
that along with $(123456)$ one should consider all the
$5!$ different cycles formed by the same $6$ elements
-- and only two of these $5!$ possibilities are shown in
the picture. To obtain our operators one should sum over all
these options. One should also add all the other cycles:
each $\Delta$ is a set of a few cycles of given lengths.

Advantage of this pictorial representation is that one
can further represent such pictures, the Feynman diagrams
by operators. This is the simplest way to obtain
(\ref{WDthrD}), which immediately reproduces eqs.(\ref{W21|3})
and (\ref{W3|21}).
The normal ordering appears because one connecting line
can not act on another connecting line.

This Feynman diagram technique ties together the geometric interpretation
of the Hurwitz numbers, their combinatorial expressions and the normal ordered
differential matrix operators.

\subsection{Algebra of cut-and-join operators}

\subsubsection{Examples of normal ordering
\label{exano}}

We begin with a few examples, illustrating role of the
normal ordering:
$$
:\tr \hat D^2:\ = \tr \hat D^2 - N\tr \hat D = \tr (\hat D-N)\hat D
$$
or
\be
\tr \hat D^2 =\ :\tr \hat D^2:\ +\ N\tr \hat D,
\nn \\
\tr \hat D^3 =\ :\tr \hat D^3:\ +\ 2N:\tr \hat D^2:\
+\ :(\tr \hat D)^2: +\ N^2\tr \hat D,
\nn\\
\tr \hat D^4 =\ :\tr \hat D^4:\ +\ 3N:\tr \hat D^3:\ +\ 3:\tr \hat
D\ \tr \hat D^2:\ +\ (3N^2+1):\tr \hat D^2:\ +\ 3N:(\tr \hat D)^2:\
+\ N^3:\tr \hat D:,
\nn \\
\ldots
\ee
Similarly,
\be
(\tr \hat D)^2 =\ :(\tr \hat D)^2:\ + \ \tr \hat D,
\ee
\be
(\tr \hat D^2)^2 =\ :(\tr \hat D^2)^2:\
+\ 2N:\tr \hat D\, \tr \hat D^2:\ + \
4:\tr \hat D^3:\ +\ 4N:\tr \hat D^2:\
+\ (N^2+2):(\tr \hat D)^2:\ +\ N^2\tr \hat D
\ee
and so on.

\subsubsection{Insertion of extra $\hat D_1$, (\ref{permid})
\label{D1}}

Next we provide a complete description of normal ordering
for a small but important class of operators which contain degrees
of $\hat D_1 = \tr \hat D
= \sum_a ap_a\frac{\p}{\p p_a}$,
\be
\hat{\cal W}([\Delta,\underbrace{1,...,1}_{k}])={1\over k!}\
:\widetilde{D(\Delta)\hat D_1^k}:
\ee
($\Delta$ is assumed not to contain more units).
For a more systematic description see \cite{ammops}.

The following relations follow directly from the
definition of normal ordering:
\be
: \widetilde{D(\Delta)} \hat D_1 :\ =\
: \widetilde{D(\Delta)}:\, \hat D_1  -
|\Delta| : \widetilde{D(\Delta)}:\  =\
: \widetilde{D(\Delta)}:\Big(\hat D_1-|\Delta|\Big),\\
: \widetilde{D(\Delta)} (\hat D_1)^2 :\ =\
: \widetilde{D(\Delta)} \hat D_1 : \hat D_1 -
(|\Delta|+1): \widetilde{D(\Delta)} \hat D_1 :\ =\
: \widetilde{D(\Delta)}\hat D_1:\Big(\hat D_1-|\Delta|-1\Big)\
= \\
=\ : \widetilde{D(\Delta)}:\Big(\hat D_1-|\Delta|\Big)
\Big(\hat D_1-|\Delta|-1\Big), \\
\ldots\\
: \widetilde{D(\Delta)} (\hat D_1)^k :\ =\ :
\widetilde{D(\Delta)}:\prod_{i=0}^{k-1} \Big(\hat D_1-|\Delta|-i\Big)
\label{ww1}\ee This
implies that when one acts with $\ : \widetilde{D(\Delta)}:\ $ on some
quantity of weight $|R|$, for example on $\ : \widetilde{D(R)}:\ $,
 then $\hat D_1$ acts as multiplication by $|R|$, and one can
always substitute $: \widetilde{D(\Delta)} :$ with
\be
\frac{1}{(|R|-|\Delta|)!} \ : \widetilde{D(\Delta)}
(\hat D_1)^{|R|-|\Delta|} :\ = \ : \widetilde{D
([\Delta,\underbrace{1,\ldots,1}_{|R|-|\Delta|}])}:\ , \ \ \ \ \ {\rm
provided}\ \ \ 1\notin \Delta \ee without changing the result, in
accordance with rule (\ref{phiext}) and with formula (\ref{permid}).

If $\hat{\cal W}(\Delta)$ contains $\hat D_1$-factors,
this rule should be modified by a numerical factor: for example
$: \widetilde{D([1])}:$ is substituted with
\be\label{ww12}
\frac{1}{(|R|-1)!} \, :
\widetilde{D([1])} (\hat D_1)^{|R|-1} :\ = \frac{1}{(|R|-1)!} \, :
(\hat D_1)^{|R|} :\ = |R|\, : \widetilde{(\hat D_1)^{|R|}} :\ =
|R|\, : \widetilde{D([\underbrace{1,\ldots,1}_{|R|}])}: \ee which contains an
extra factor of $|R|$, again in accordance with (\ref{phiext}).

\subsubsection{Multiplication algebra of $\hat{\cal W}$-operators
\label{Wal}}

Making use of relations from s.\ref{exano} we can now multiply
different cut-and-join operators:
\be
\boxed{
\hat{\cal W}(\Delta_1)\hat{\cal W}(\Delta_2) =
\sum_\Delta C_{\Delta_1\Delta_2}^\Delta \hat{\cal W}(\Delta)
}
\label{CWW}
\ee
Note that in variance with (\ref{Ccc}) there is no restriction
on the sizes of Young diagrams $\Delta_1,\Delta_2$ and $\Delta$,
actually, there is only a selection rule
\be
{\rm max} (|\Delta_1|,|\Delta_2|)\leq |\Delta| \leq
|\Delta_1| + |\Delta_2|
\ee
Still these new structure constants with
$|\Delta_1|=|\Delta_2|=|\Delta|$ coincide with the structure
constants of conjugation-classes algebra (\ref{Ccc}). The Feynman
diagram technique of s.\ref{dia} can be considered as
a pictorial representation of (\ref{CWW}), while expression
(\ref{Wpp}) for the $\hat{\cal W}$ operator through
the time-variables -- as a corollary of
(\ref{CWW}) projected to the $|\Delta_1|=|\Delta_2|=|\Delta|$
subset. In this case it implies that
\be
\hat{\cal W}(\Delta_1)\widetilde{p\,(\Delta_2)} =
\sum_\Delta C_{\Delta_1,\Delta_2}^\Delta \widetilde{p\,(\Delta)}
= \hat{\cal W}(\Delta_2)\widetilde{p\,(\Delta_1)},
\ \ \ \ \ \ |\Delta_1|=|\Delta_2|
\ee

Furthermore, in accordance with (\ref{Wchi}) the eigenvalues
$\varphi_R(\Delta)$ satisfy the same algebra (\ref{CWW}):
\be
\boxed{
\varphi_R(\Delta_1)\varphi_R(\Delta_2) =
\sum_\Delta C_{\Delta_1\Delta_2}^\Delta
\varphi_R(\Delta)
}
\label{Cpp}
\ee
The structure constants in this relation do not depend on $R$,
which is not so obvious if one extracts $\varphi_R(\Delta)$ from the character
expansion (\ref{varphichi}). One can explicitly check this for
first few $\varphi_R(\Delta)$ using the following table for them ($\varphi_R(\Delta)$
differs by a factor from the character of symmetric group \cite{Mac})):

\bigskip

\centerline{
$
\begin{array}{|c||c|cc|ccc|ccccc|ccccccc|}
\hline
&&&&&&&&&&& &&&&&&&\\
R\diagdown \Delta&1 &2&11 &3&21&111 &4&31&22&211&1111 &5&41&32&311&221&2111&11111\\
&&&&&&&&&&& &&&&&&&\\
\hline \hline
&&&&&&&&&&& &&&&&&&\\
1&1&&&&&&&&&& &&&&&&&\\
&&&&&&&&&&& &&&&&&&\\
\hline
&&&&&&&&&&& &&&&&&&\\
2&2&1&1&&&&&&&& &&&&&&&\\
&&&&&&&&&&& &&&&&&&\\
{11}&2&-1&1&&&&&&&& &&&&&&&\\
&&&&&&&&&&& &&&&&&&\\
\hline
&&&&&&&&&&& &&&&&&&\\
3&3&3&3&2&3&1&&&&& &&&&&&&\\
&&&&&&&&&&& &&&&&&&\\
{21}&3&0&3&-1&0&1&&&&& &&&&&&&\\
&&&&&&&&&&& &&&&&&&\\
{111}&3&-3&3&2&-3&1&&&&& &&&&&&&\\
&&&&&&&&&&& &&&&&&&\\
\hline
&&&&&&&&&&& &&&&&&&\\
4&4&6&6&8&12&4&6&8&3&6&1 &&&&&&&\\
&&&&&&&&&&& &&&&&&&\\
{31}&4&2&6&0&4&4&-2&0&-1&2&1 &&&&&&&\\
&&&&&&&&&&& &&&&&&&\\
{22}&4&0&6&-4&0&4&0&-4&3&0&1 &&&&&&&\\
&&&&&&&&&&& &&&&&&&\\
{211}&4&-2&6&0&-4&4&2&0&-1&-2&1 &&&&&&&\\
&&&&&&&&&&& &&&&&&&\\
{1111}&4&-6&6&8&-12&4&-6&8&3&-6&1 &&&&&&&\\
&&&&&&&&&&& &&&&&&&\\
\hline
&&&&&&&&&&& &&&&&&&\\
5&5&10&10&20&30&10&30&40&15&30&5 &24&30&20&20&15&10&1\\
&&&&&&&&&&& &&&&&&&\\
{41}&5&5&10&5&15&10&0&10&0&15&5 &-6&0&-5&5&0&5&1\\
&&&&&&&&&&& &&&&&&&\\
{32}&5&2&10&-4&6&10&-6&-8&3&6&5 &0&-6&4&-4&3&2&1\\
&&&&&&&&&&& &&&&&&&\\
{311}&5&0&10&0&0&10&0&0&-5&0&5 &4&0&0&0&-5&0&1\\
&&&&&&&&&&& &&&&&&&\\
{221}&5&-2&10&-4&-6&10&6&-8&3&-6&5 &0&6&-4&-4&3&-2&1\\
&&&&&&&&&&& &&&&&&&\\
{2111}&5&-5&10&5&-15&10&0&10&0&-15&5 &-6&0&5&5&0&-5&1\\
&&&&&&&&&&& &&&&&&&\\
{11111}&5&-10&10&20&-30&10&-30&40&15&-30&5 &24&-30&-20&20&15&-10&1\\
&&&&&&&&&&& &&&&&&&\\
\hline
\end{array}
$
}

\subsubsection{Examples of structure constants}

We now give some explicit examples of (\ref{CWW}):
a multiplication table
restricted to the case when $|\Delta|\leq 4$.
Many of them are direct corollaries of relations
from s.\ref{D1}.
Note that explicit $N$-dependence which showed up
in the normal-ordered products in s.\ref{exano}
drops away when one considers products of the normal-ordered
operators $\hat {\cal W}$.

Underlined are the components
satisfying $|\Delta_1|=|\Delta_2|=|\Delta|$, which are
dictated by compositions of permutations, eq.(\ref{Ccc}):
$$
\underline{\hat{\cal W}([1])\hat{\cal W}([1])} =
\underline{\hat{\cal W}([1])} + 2\hat{\cal W}([1,1]),
$$

$$
\hat{\cal W}([1])\hat{\cal W}([2]) = 2\hat{\cal W}([2]) + \hat{\cal W}([2,1]),
$$
$$
\hat{\cal W}([1])\hat{\cal W}([1,1]) = 2\hat{\cal W}([1,1]) + 3\hat{\cal W}([1,1,1]),
$$

$$
\hat{\cal W}([1])\hat{\cal W}([3]) = 3\hat{\cal W}([3]) + \hat{\cal W}([3,1]),
$$
$$
\hat{\cal W}([1])\hat{\cal W}([2,1]) = 3\hat{\cal W}([2,1]) + 2\hat{\cal W}([2,1,1]),
$$
$$
\hat{\cal W}([1])\hat{\cal W}([1,1,1])
= 3\hat{\cal W}([1,1,1]) + 4\hat{\cal W}([1,1,1,1]),
$$

$$
\hat{\cal W}([1])\hat{\cal W}([4]) = 4\hat{\cal W}([4]) + \hat{\cal W}([4,1]),
$$
$$
\hat{\cal W}([1])\hat{\cal W}([3,1]) = 4\hat{\cal W}([3,1]) + 2\hat{\cal W}([3,1,1]),
$$
$$
\hat{\cal W}([1])\hat{\cal W}([2,2]) = 4\hat{\cal W}([2,2]) + \hat{\cal W}([2,2,1]),
$$
$$
\hat{\cal W}([1])\hat{\cal W}([2,1,1])
= 4\hat{\cal W}([2,1,1]) + 3\hat{\cal W}([2,1,1,1]),
$$
$$
\hat{\cal W}([1])\hat{\cal W}([1,1,1,1])
= 4\hat{\cal W}([1,1,1,1]) + 5\hat{\cal W}([1,1,1,1,1]),
$$

$$
\underline{\hat{\cal W}([1,1])\hat{\cal W}([2])} =
\underline{\hat{\cal W}([2])} + 2\hat{\cal W}([2,1]) + \hat{\cal W}([2,1,1]),
$$
$$
\underline{\hat{\cal W}([1,1])\hat{\cal W}([1,1])} =
\underline{\hat{\cal W}([1,1])} + 6\hat{\cal W}([1,1,1]) + 6\hat{\cal W}([1,1,1,1]),
$$
$$
\underline{\hat{\cal W}([2])\hat{\cal W}([2])} =
\underline{\hat{\cal W}([1,1])} + 3\hat{\cal W}([3]) + 2\hat{\cal W}([2,2]),
$$

$$
\hat{\cal W}([1,1])\hat{\cal W}([3]) =
3\hat{\cal W}([3]) + 3\hat{\cal W}([3,1]) + \hat{\cal W}([3,1,1]),
$$
$$
\hat{\cal W}([1,1])\hat{\cal W}([2,1]) =
3\hat{\cal W}([2,1]) + 6\hat{\cal W}([2,1,1]) + \hat{\cal W}([2,1,1,1]),
$$
$$
\hat{\cal W}([1,1])\hat{\cal W}([1,1,1]) =
3\hat{\cal W}([1,1,1]) + 12\hat{\cal W}([1,1,1,1]) + 10\hat{\cal W}([1,1,1,1,1]),
$$
$$
\hat{\cal W}([2])\hat{\cal W}([3]) = \hat{\cal W}([3,2])+4\hat{\cal
W}([4])+2\hat{\cal W}([2,1])
$$
$$
\hat{\cal W}([2])\hat{\cal W}([2,1]) = 2\hat{\cal W}([2,2,1])+3\hat{\cal
W}([3,1])+4\hat{\cal W}([2,2])+3\hat{\cal W}([3])+3\hat{\cal W}([1,1,1])
$$
$$
\hat{\cal W}([2])\hat{\cal W}([1,1,1]) =
\hat{\cal W}([2,1]) + 2\hat{\cal W}([2,1,1]) + \hat{\cal W}([2,1,1,1]),
$$

$$
\ldots
$$

\subsection{From $\hat D$ to differential
operators in time-variables\label{Winp}}

One way to express the operators $\hat{\cal W}$ through
the time-variables is already given in eq.(\ref{WDthrD}).
However, it is much simpler to extract such expressions
directly from (\ref{DPW}), i.e. by making a Miwa transformation
back from the matrix-$X$ variable to times $p_k = \tr X^k$.
This is done by the simple rule: when acting on a function
of time-variables, the $X$-derivatives provide
\be
\hat D_{ab} F(p) = X_{ac}\frac{\p}{\p X_{bc}} F(p) =
\sum_{k=1}^\infty k(X^k)_{ab} \frac{\p F(p)}{\p p_k}
\ee
Next $\hat D$ operators act both on $X$ which
emerged at the first stage and on the remaining function
of time-variables:
\be
\hat D_{a'b'}\hat D_{ab} F(p) = \sum_{k,l=1}^\infty
kl(X^l)_{a'b'}(X^k)_{ab}\frac{\p^2F(p)}{\p p_k p_l} +
\sum_{k=1}^\infty\sum_{j=0}^{k-1}k (X^j)_{ab'}(X^{k-j})_{a'b} \frac{\p F(p)}{\p p_k}
\label{DDF}
\ee
where we used the fact that
\be
\hat D_{a'b'}(X^k)_{ab} =
X_{a'c'}\frac{\p}{\p X_{b'c'}}(X^k)_{ab}
= \sum_{j=0}^{k-1} X_{a'c'} (X^j)_{ab'}(X^{k-j-1})_{c'b}
= \sum_{j=0}^{k-1} (X^j)_{ab'} (X^{k-j})_{a'b}
\ee
Note that the power of $X$ in the second factor at the r.h.s.
is always non-vanishing, while it can vanish in the first
factor. If we considered a normal ordered product
of operators instead of (\ref{DDF}), this power would also
be non-vanishing:
\be
:\hat D_{a'b'}\hat D_{ab}:\, F(p) =
\sum_k\left(k\sum_{j=1}^{k-1} (X^j)_{ab'} (X^{k-j})_{a'b}\right) {\p F(p)\over\p p_k}+
\sum_{k,l} kl(X^k)_{ab}(X^l)_{a'b'}{\p ^2 F(p)\over\p p_k\p p_l}
\ee
This is the property that guarantees the potential $N$-dependence
is eliminated from the formulas, as it should be for the
operators expressible in terms of time-variables,
and thus independent of details of the Miwa transform
(of which $N$ is an additional parameter).

First few examples of the cut-and-join operators in terms of the time-variables are
\be
\hat{\cal W}([1]) = \tr \hat D= \sum_{k=1} kp_k\frac{\p}{\p p_k}
\ee
\be
\hat{\cal W}([2]) ={1\over 2} \, :\tr \hat D^2\,:\ =
\frac{1}{2}\sum_{a,b=1}^\infty
\left( (a+b)p_ap_b\frac{\p}{\p p_{a+b}} +
abp_{a+b}\frac{\p^2}{\p p_a\p p_b}\right)
\ee
\be
\hat{\cal W}([1,1]) = \frac{1}{2!}\, :(\tr \hat D)^2\,:\ =
\frac{1}{2}\left(\sum_{a=1}^\infty
 a(a-1)p_a\frac{\p}{\p p_{a}} +\sum_{a,b=1}^\infty
abp_{a}p_b\frac{\p^2}{\p p_a\p p_b}\right)
%= \frac{1}{2}\hat{\cal W}([1])\Big(\hat{\cal W}([1])-1\Big)
\ee
\be
\hat{\cal W}([3]) = \frac{1}{3}\, :\tr \hat D^3\,:\ =
\frac{1}{3}\sum_{a,b,c\geq 1}^\infty
abcp_{a+b+c} \frac{\p^3}{\p p_a\p p_b\p p_c}
+ \frac{1}{2}\sum_{a+b=c+d} cd\left(1-\delta_{ac}\delta_{bd}\right)
p_ap_b\frac{\p^2}{\p p_c\p p_d} + \\
+ \frac{1}{3} \sum_{a,b,c\geq 1}
(a+b+c)\left(p_ap_bp_c + p_{a+b+c}\right)\frac{\p}{\p p_{a+b+c}}
\label{W3p}
\ee
\be
\hat{\cal W}([2,1]) = \frac{1}{2}\, :\tr \hat D^2\,\tr \hat D \,:\ =
{1\over 2}\sum_{a,b\ge 1}(a+b)(a+b-2)p_ap_{b}{\p\over\p p_{a+b}}\,+
{1\over 2}\sum_{a,b\ge 1}ab(a+b-2)p_{a+b}{\p ^2\over\p p_a\p p_b}\,+
\nn \ee \be %\\
+\frac{1}{2}\sum_{a,b,c\ge 1} (a+b)cp_ap_bp_c\frac{\p^2}{\p p_{a+b}\p p_c}
%%%+\frac{1}{4}\sum_{a+b=c+d} cd\left(1-\delta_{ac}\delta_{bd}\right)
%%%p_ap_b\frac{\p^2}{\p p_c\p p_d}
+{1\over 2}\sum_{a,b,c\ge 1}abcp_ap_{b+c}{\p ^3\over\p p_a\p p_b\p p_c}
\ee
\be
\hat{\cal W}([1,1,1]) = \frac{1}{3!}\, :(\tr \hat D)^3\,:\ =
{1\over 6}\sum_{a\ge 1} a(a-1)(a-2)p_a{\p\over\p p_a}\,+\\
%%%+{1\over 2}\sum_{{a,b,c,d\ge 1}\atop{a+b=c+d}}abp_cp_{d}{\p ^2\over\p p_a\p p_b}
+ {1\over 4}\sum_{a,b} ab(a+b-2)p_ap_b\frac{\p^2}{\p p_a\p p_b}\,
+{1\over 6}\sum_{a,b,c\ge 1}abcp_ap_bp_c{\p ^3\over\p p_a\p p_b\p p_c}
%= \frac{1}{6}\hat{\cal W}([1])\Big(\hat{\cal W}([1])-1\Big)\Big(\hat{\cal W}([1])-2\Big)
\ee
As one had to expect from (\ref{ww1}) and (\ref{ww12}), it follows from these formulas that
\be
\hat{\cal W}([1,1])={1\over 2}\hat{\cal W}([1])(\hat{\cal W}([1])-1)\\
\hat{\cal W}([2,1])=\hat{\cal W}([2])(\hat{\cal W}([1])-2)\\
\hat{\cal W}([1,1,1])={1\over 6}\hat{\cal W}([1])(\hat{\cal W}([1])-2)(\hat{\cal W}([1])-1)
\ee

The manifest expressions for higher operators fast become much more involved. However,
there is a much more compact presentation for the operators:
when expressed through the time-variables, operators are in fact
made from the scalar field current
\be
\p\Phi(z) = \sum_k \left(kt_kz^k + \frac{1}{z^k}\frac{\p}{\p t_k}\right)=
\sum_k \left(p_kz^k + \frac{k}{z^k}\frac{\partial}{\partial p_k}\right)
\ee
and from an additional dilatation operator
\be
\hat R = \left(z\frac{\p}{\p z}\right)^2
\ee
For more details see \cite{integ,ammops}. Here we provide
just a few simplest examples.

The normal ordering in \textit{these} formulas means that all
$p$ factors stand to the left of $\p/\p p$ factors,
we do not take $p$-derivatives of $p$'s when building
up an operator from $\p\Phi(z)$.
The subscript $0$ means that one
should pick up the coefficient in front of $z^0$ in the
$z$-series. As soon as adding units to the Young diagram is a trivial procedure, as we
just saw, we list here only the operators corresponding to the Young diagrams
without units \cite{ammops}:
\be
\hat{\cal W}([1])=\hat {{C}}_{1}\\
\hat{\cal W}([2])={1\over 2}\hat {{C}}_{2}\\
\hat{\cal W}([3])={1\over 3}\hat {{C}}_{3}-{1\over 2}\hat {{C}}_{1}^2+{1\over 3}
\hat {{C}}_{1}\\
\hat{\cal W}([2,2])={1\over 8}\hat {{C}}_{2}^2-{1\over 2}\hat {{C}}_{3}
+{1\over 2}\hat {{C}}_{1}^2-{1\over 4}\hat {{C}}_{1}\\
\hat{\cal W}([4])={1\over 4}\hat {{C}}_{4}-\hat {{C}}_{1}
\hat {{C}}_{2}+{5\over 4}\hat {{C}}_{1}
\ee
where the Casimir operators are \cite{ammops}
\be
\hat {{C}}_{1}=\frac{1}{2}:\left[(\p \Phi)^2\right]_0:\nn\\
\hat {{C}}_{2}=\frac{1}{3}:\left[(\p \Phi)^3\right]_0:\nn\\
\hat {{C}}_{3}=\frac{1}{4}:\left[(\p \Phi)^4+\p \Phi(\hat R\p \Phi)\right]_0:\nn\\
\hat {{C}}_{4}=\frac{1}{5}:\left[(\p \Phi)^5+\frac{5}{2}(\p \Phi)^2(\hat R\p \Phi)\right]_0:
\ee
\be
\hat {{C}}_{k}=\frac{1}{k+1}:\left[(\p \Phi)^{k+1}+\frac{(k+1)!}{4!(k-2)!}(\p \Phi)^{k-1}(\hat R\p \Phi)+\ldots\right]_0:\nn\\
\ee

\subsection{$GL(\infty)$ characters and related formulas
\label{chars}}

$GL$ characters $\chi_R(t)$ are defined with the help of
the first Weyl determinant formula
\be
\chi_R(t) = \det_{ij} s_{\mu_i+j-i}(t)
\ee
where $s_k(t)$ are the Shur polynomials,
\be
\exp\left(\sum_k t_kz^k\right) = \sum_k s_k(t)z^k
\ee
After the Miwa transformation $p_k=kt_k=\tr X^k$,
the same characters are expressed through the eigenvalues
of matrix $X$ by the second Weyl formula
\be
\chi_R[X] = \chi_R\Big(t_k = \frac{1}{k}\tr X^k\Big)
= \frac{\det_{ij} x_i^{\mu_j-j}}{\det_{ij} x_i^{-j}}
\ee
The expansion of $\chi_R(t)$ in powers of $p$'s defines
the coefficients $\varphi_R(\Delta)$ by
eq.(\ref{varphichi}) for $|R|=|\Delta|$ and by eq.(\ref{phiext})
for all other $\Delta$'s. In (\ref{varphichi}) the parameter
$d_R$ is the value of character at the point
$t_k = \delta_{k,1}$,
\be
d_R = \chi_R(\delta_{k1})
\ee
and it is given by the hook formula
\be
d_R = \prod_{{\rm all\ boxes\ of}\ R}\
\frac{1}{{\rm hook\ length}}
= \dfrac{ \prod\limits_{i < j = 1}^{|R|}
\left( \mu_i - \mu_j - i + j \right) }
{\prod\limits_{i = 1}^{|R|} \left(\mu_i + |R| - i\right)!}
\ee
One can also introduce a natural scalar product on the characters
\be
\left<\chi_R(t),\chi_{R'}(t)\right>=\delta_{RR'}
\ee
manifestly given by the formula
\be
\left.\left<A(t),B(t)\right>\equiv A\left({\partial\over \partial p}\right)B(t)\right|_{t_k=0}
\ee
In particular,
\be
\left<p(\Delta),\widetilde{p(\Delta')}\right>=\delta_{\Delta\Delta'}
\ee
These formulas, along with (\ref{varphichi}) immediately lead to the inverse expansion
\be\label{pchi}
\widetilde{p(\Delta)}
= \sum_{R} d_R\varphi_R(\Delta)\chi_R(t)
\delta_{|\Delta|,|R|}
\ee

Thus (\ref{Wchi}) is actually an exhaustive alternative
definition of the operators $\hat {\cal W}(\Delta)$
and one can check that (\ref{WDthrD})
does satisfy this definition (see the next subsection s.2.7 and \cite{integ,ammops}).
The equivalence of two definitions (\ref{DPW})
and (\ref{Wchi}) follows from formula
(\ref{Frofor}).

\subsection{From (\ref{WDthrD}) to (\ref{Wchi})\label{mf}}

The idea of a straightforward derivation of (\ref{Wchi})
from (\ref{WDthrD}) goes as follows. For the sake of simplicity consider $\Delta$ that
does not contain units (the generalization is straightforward).
Obviously,
\be
:\widetilde{D(\Delta)}: e^{t_1} =
:\widetilde{D(\Delta)}: e^{\tr X}
= \widetilde{p(\Delta)} e^{t_1}
\ee
Since
\be\label{ff}
e^{t_1} = \sum_R d_R\chi_R(t)
\ee
it follows that
\be
\sum_R d_R :\widetilde{D(\Delta)}:\chi_R(t) =
\widetilde{p(\Delta)}e^{t_1}
\label{dDc}
\ee
The r.h.s. of this formula can be rewritten using (\ref{pchi}) and (\ref{phiext}) as
\be\label{pphi}
\sum_k\widetilde{p(\Delta)}{t_1^k\over k!}=\sum_k \widetilde{p([\Delta,\underbrace{1,...,1}_k])}=
\sum_k\sum_R d_R\varphi_R([\Delta,\underbrace{1,...,1}_k])\chi_R=\\=
\sum_k\sum_{R:\ |R|=|\Delta|+k}d_R\varphi_R(\Delta)\chi_R=\sum_R d_R\varphi_R(\Delta)\chi_R
\ee
which, along with (\ref{dDc}) and the fact that $\chi_R(t)$ are the eigenfunctions of
$:\widetilde{D(\Delta)}:\ $, ultimately leads to (\ref{Wchi}).

\subsection{Details of (\ref{Ztt})
\label{Zttsec}}

Deviation from the naive formula (\ref{Ztt})
arises because one should carefully impose
the condition $|\Delta|=R$ in (\ref{varphichi})
when passing from $\varphi_R(\Delta)$ in
(\ref{Frofor}) to the characters $\chi_R(t')$
in (\ref{calZ}):
\be
Z(t,t',\ldots) = \sum_q \left\{\sum_{\Delta,\Delta'}
p(\Delta)p'(\Delta')\delta_{|\Delta|,q}\delta_{|\Delta'|,q}
 {\rm Cov}_q(\Delta,\Delta',\ldots)\right\}
= \sum_{R}\chi_R(t)\chi_R(t')\ldots
\ee
or, alternatively
\be
= \sum_{\Delta',R} d_R\chi_R(t) \varphi_R(\Delta)p'(\Delta)\ldots
\delta_{|\Delta|,|R|}
= \oint \frac{dz}{z}
\sum_{\Delta,R}d_R\chi_R(t) \varphi_R(\Delta)p'(\Delta)z^{|\Delta|-|R|}\ldots
= \nn \\
= \oint \frac{dz}{z} \sum_\Delta z^{|\Delta|}p'(\Delta):\widetilde{D(\Delta)}:
\sum_R d_R\chi_R(t)z^{|R|}\ldots =
\oint \frac{dz}{z}\ :\exp\left(\sum_{k=1}^\infty z^{k}t'_k\hat D_k\right):\
\sum_R d_R\chi_R(t)z^{|R|}\ldots
\ee
This is the full (correct) version of eq.(\ref{Ztt}).
If there is nothing at the place of dots at the r.h.s., i.e. if one considers
$Z(t,t'|0)$, then the sum over $R$ is equal to $e^{t_1/z}$ and one obtains,
as a simplest example,
\be
Z(t,t'|0) =
\oint \frac{dz}{z}
\ :\exp\left(\sum_{k=1}^\infty z^{k}t'_k\hat D_k\right):\
e^{t_1/z} = \exp\left(\sum_k kt_kt'_k\right)
\oint \frac{dz}{z} e^{t_1/z} = \exp\left(\sum_k kt_kt'_k\right)
\ee
Generalizations are straightforward.
If one wants to consider multiple Hurwitz numbers, with more sets
of $t$-variables, then extra $\delta$-functions and $z$-integrals
should be included.

\section{Summary and conclusion}

The generic cut-and-join operator $\hat{\cal W}(\Delta)$
is associated with the arbitrary Young diagram $\Delta$ and
can be defined in two alternative ways.

First, through the characters, requesting that
\be
\hat{\cal W}(\Delta) \chi_R(t) = \varphi_R(\Delta) \chi_R(t)
\label{Ochi}
\ee
for any Young diagram $R$.
Then the Hurwitz partition function, say for two sets
time-variables, is equal to
\be
{\cal Z}(t,\bar t|\beta) \equiv \sum_R \chi_R(t)\chi_R(\bar t)
\exp \left(\sum_\Delta \beta_\Delta\varphi_R(\Delta)\right)
= \exp\left(\sum_\Delta \beta_\Delta \hat{\cal W}(\Delta)\right)
Z(t,\bar t|0)
\ee
where $Z(t,\bar t|0) = \sum_R \chi_R(t)\chi_R(\bar t)
= \exp\left(\sum_k kt_k\bar t_k\right)
= \exp\left(\sum_k \frac{1}{k}p_k\bar p_k\right)$.
The Hurwitz Toda $\tau$-function \cite{KM2,OkToda} for
"the double Hurwitz numbers",
$Z(t,\bar t|\beta) = \sum_R \chi_R(t)\chi_R(\bar t)
e^{\beta_2\varphi_R([2])} = e^{\beta_2\hat{\cal W}([2])}Z(t,\bar t,0)$,
is a particular case,
and the Kontsevich-Hurwitz KP $\tau$-function for
"the single (or ordinary) Hurwitz numbers" is a further
restriction to $\bar t_k = \delta_{k,1}$.
Integrability in these two examples is preserved
because the simplest cut-and-join operator
$\hat{\cal W}([2])$ coincides with the (second) Casimir operator,
and integrability is present when any combination of Casimir
operators acts on the $\tau$-function \cite{KM2}.
It is violated by the generic cut-and-join operators with
$|\Delta|\geq 3$.
Of course, the $\beta$-variables can be considered as
associated with some \textit{new} integrable structure,
reflected in commutativity of the operators $\hat{\cal W}(\Delta)$,
\be
\left[\hat{\cal W}(\Delta_1), \hat{\cal W}(\Delta_2)\right] = 0
\ \ \ \ \forall \ \Delta_1 \ {\rm and}\  \Delta_2
\ee
However, there is no obvious way to relate \textit{this}
integrability to the group-theory-induced Hirota-like bilinear
relations \cite{gentau}, and there are even less chances that
it is somehow induced by the free-fermion representation of
$\widehat{U(1)}$ (these are the two features built into the
definition of integrable hierarchies of the KP/Toda type).

Second, through permutations and their cyclic decompositions.
This is the problem directly related to merging of ramification
points of the Hurwitz covering.
The central formula is (\ref{DPW}),
\be
\Delta \circ {\cal P}(p) = \hat{\cal W}(\Delta){\cal P}(p)
\label{Operc}
\ee
and in s.\ref{dpw} we explained how $\hat{\cal W}(\Delta)$
is explicitly reconstructed from the knowledge of
permutation compositions.

Both these definitions, conditions (\ref{Ochi}) and
(\ref{Operc}) are explicitly resolved by eq.(\ref{WDthrD}),
\be
\hat {\cal W}(\Delta) = \ :\widetilde{D(\Delta)}:
\ee
-- a direct generalization of representation for the simplest
cut-and-join operator $\hat{\cal W}([2])$ from \cite{MShW}.
The first few operators of this set are listed in sect.\ref{Winp}.

These operators form a commutative associative algebra
\be
\hat{\cal W}(\Delta_1)\hat{\cal W}(\Delta_2) =
\sum_\Delta C_{\Delta_1\Delta_2}^\Delta \hat{\cal W}(\Delta)
\ee
We call it Universal Hurwitz Algebra, because it
allows one to define the Universal Hurwitz Numbers
for arbitrary collection of Young diagrams, not
obligatory of the same size. That is, if complemented
by (43), formula (\ref{Frofor}) allows
one to define the Hurwitz number as
\be
{\rm Cov}(\Delta_1,\ldots,\Delta_m) =
\sum_\Delta C_{\Delta_1\ldots\Delta_m}^\Delta
\Big(\sum_R d_R^2 \varphi(\Delta)\Big)
\ee
where $C$ is a combination of structure constants,
for example,
\be
C_{\Delta_1\ldots\Delta_m}^\Delta =
\sum_{\Delta_a,\Delta_b,\ldots,\Delta_c}
C_{\Delta_1\Delta_2}^{\Delta_a}
C_{\Delta_a\Delta_3}^{\Delta_b}
\ldots
C_{\Delta_c\Delta_m}^\Delta
\ee
(the order of pairing is actually inessential
because of associativity and commutativity of the
algebra).
Thus, evaluation of the Hurwitz numbers is reduced to
evaluation of the single form on the linear algebra
of Young diagrams
\be
\sum_R d^2_R\varphi_R(\Delta) =
\left\{\begin{array}{cc}
0 & {\rm if}\ \Delta\ {\rm contains\ more\ than\ one\ column}\\
\frac{1}{n!} & {\rm if}\ \Delta = [\underbrace{1,\ldots,1}_{n}]
\end{array}
\right.
\ee
This fact that the form is vanishing except on the
single-column diagrams is a direct corollary of the
fundamental sum rule (\ref{ff}).

More details about the character-related description and
integrability aspects of the problem can be found in
accompanying papers \cite{integ,ammops}.
All these relations deserve better and more explicit
understanding.
Of special interest is study of the matrix-model representation
\cite{MShW} for the Hurwitz KP $\tau$-function,
its application (using \cite{AMMIM}) to understanding of the mysterious relation of twisting
of the Hurwitz KP $\tau$-function
to the Kontsevich $\tau$-function \cite{giv,kaz,mmhk}
and further non-Abelian generalization of this entire formalism to
the Hurwitz numbers for coverings of Riemann surfaces
with boundaries \cite{AN} (e.g. disk instead of Riemann sphere).

\section*{Acknowledgements}

We are indebted to Sasha Alexandrov and Shamil Shakirov
for useful discussions and comments.

Our work is partly supported by Russian Federal Nuclear Energy
Agency, by the joint grants 09-02-91005-ANF, 09-02-90493-Ukr and 09-01-92440-CE,
by the Russian President's Grants of
Support for the Scientific Schools NSh-3035.2008.2
(A.M.'s) and NSh-709.2008.1 (S.N.), by RFBR grants
07-02-00878 (A.Mir.), 07-02-00645
(A.Mor.) and 07-01-00593 (S.N.).

\end{document}